\documentclass[11pt]{article}

\usepackage[margin=1in]{geometry}
\usepackage[utf8]{inputenc}
\usepackage[T1]{fontenc}
\usepackage{lmodern}
\usepackage{amsmath,amssymb}
\usepackage{graphicx}
\usepackage{booktabs}
\usepackage{array}
\usepackage{longtable}
\usepackage{xcolor}
\usepackage{fancyvrb}
\usepackage[hidelinks]{hyperref}
\usepackage{authblk}
\usepackage{caption}

\providecommand{\Description}[1]{}

\title{\textbf{bFaaaP: An Inclusive, Head-Angle Piano-Pedal Interaction that
Quantitatively Reproduces a Pianist's Intended Pedalling---Foot-Free, for
Acoustic and Electronic Pianos}}

\author[1,2]{Tomoyuki Shishido\thanks{Corresponding author and first author.
ORCID: \href{https://orcid.org/0000-0002-8944-2088}{0000-0002-8944-2088}
(\url{https://orcid.org/0000-0002-8944-2088}). Email:
\texttt{shishid@saaipf.com}.}}
\author[3]{Masahiro Ootaki\thanks{ORCID:
\href{https://orcid.org/0009-0003-2760-2861}{0009-0003-2760-2861}.}}
\author[4]{Hiroyuki Narusawa\thanks{ORCID:
\href{https://orcid.org/0009-0005-3630-1408}{0009-0005-3630-1408}.}}
\affil[1]{Shishido \& Associates, Horinouchi, Hachioji, 192-0355, Tokyo, Japan}
\affil[2]{SeemeData Labs, Inc., 2-19-2 3F Nakano, Nakano-ku, 164-0001, Tokyo, Japan}
\affil[3]{Ootaki \& Architects, Fujimi 1-19-15, Sayama, 350-1306, Saitama, Japan}
\affil[4]{Naru Science Software, Fujimi-shi, 354-0024, Saitama, Japan}

\date{Preprint --- July 2026}

\begin{document}
\maketitle

\begin{abstract}
Expressive piano performance depends on the sustain (damper) pedal, operated by
foot---excluding players who cannot readily use their feet: wheelchair users and
others with lower-limb impairments, small children, and some elderly or disabled
players. We present \textbf{bFaaaP} (\emph{barrier-Free assist as a Pedal}), an
inclusive, foot-free interaction that operates the pedal from the angle of the
player's \emph{head}: a smartphone tracks head pose with on-device augmented-reality
(AR) face tracking and streams a compact command over Bluetooth Low Energy (BLE) to
a pedal device. Supported by patent examination, our central claim is \emph{not} the
head$\rightarrow$pedal architecture (anticipated by prior art) but a quantitative,
user-tunable control law---the patentable ``key'' to a natural, expressive result:
the player presets a small angular dead-zone (offset $3$--$10^\circ$) and a
multiplier ($10$--$50$), which together fix a secondary, pre-adjustable response
speed that reproduces the pianist's intended pedalling. An engineering trick
decouples the fast AR sampling from the slower BLE rate. Two co-equal realizations
share one controller: a non-destructive robotic actuator for acoustic pianos
(\emph{Pro}), anchored by a pneumatic ``airback'' (our coined term for an inflatable
air-braced anchor) that absorbs the reaction force without modifying the instrument;
and an electronic sustain switch for digital pianos (\emph{Switch}). In a
human-subject Auxiliary Pedal Effect Evaluation (APEE) with 15 participants, bFaaaP
significantly increased sustained-tone energy ($p<0.01$) and was statistically
indistinguishable from a player's own foot ($p>0.05$), with no significant difference
across classes; one participant with a leg disability and a tracheostomy performed
successfully. With nothing worn on the face and fast setup, bFaaaP has run in formal
public concerts (2018--2025). We release the full hardware and software as open
source.
\end{abstract}


% =====================================================================
% Paper body (Introduction -> Conclusion), included by main.tex (arXiv, article).
% Also shared with the private ACM TACCESS (acmart) version kept outside this repo.
% Figures use \Description{} (acmart defines it; article no-ops it via
% \providecommand).

\section{Introduction}
\label{sec:intro}
Expressive piano performance relies heavily on the \emph{sustain (damper) pedal},
which is operated by the foot. This foot requirement excludes a population that
\emph{can} play the keys but cannot operate the pedal: wheelchair users and others
with lower-limb impairments, small children whose feet do not yet reach the
pedals, and some elderly and severely disabled players. The project began in 2018
from exactly such a request: a wheelchair-using client asked whether the piano
pedal could be motorized so that it could be ``pressed.''

Crucially, the goal is not merely to switch a note on and off, but to let a player
produce their \emph{own intended pedalling}---the timing and depth that expressive
performance requires. We therefore frame bFaaaP as an \emph{inclusive design} and
a \emph{human--machine interaction}: a smartphone controller that senses head
angle with AI face tracking, paired with two device families that realize the
pedal action---one for acoustic pianos and one for digital instruments. Accessible
and assistive musical instruments~\cite{frid2019,ramirez2023,duarte2023,eyeharp},
hands-free head-based interfaces~\cite{sumak2019,varona2008,manresa2014,arkit}, and
robotic musical actuation~\cite{weinberg2020,scimeca2020,zappi2012} each inform the
design, and existing assistive pedal aids~\cite{canassist,steingraeber,mouthpedal2008}
show the need; bFaaaP differs in \emph{how} it reproduces intended pedalling and in
\emph{evidence} that it does so.

The contributions of this paper are:
\begin{enumerate}
  \item \textbf{An inclusive, foot-free head-angle interaction} that reproduces a
  pianist's intended pedalling, delivered by one smartphone controller and two
  co-equal device realizations (acoustic \emph{Pro}, electronic \emph{Switch}).
  \item \textbf{The control \emph{law}, not the architecture, as the novelty}: a
  quantitative, user-tunable mapping---angular dead-zone (offset $3$--$10^\circ$),
  user multiplier ($10$--$50$), full action within $+2$--$10^\circ$, and the
  pre-adjustable response speed these presets yield---rather than the bare
  head$\rightarrow$pedal chain,
  which prior art anticipates (Section~\ref{sec:control-law}).
  \item \textbf{Human-subject clinical-style validation} (APEE, 15 participants):
  bFaaaP increases sustained-tone energy ($p<0.01$), is indistinguishable from a
  player's own foot ($p>0.05$), and works equally across adults, children, and
  people with disabilities (Section~\ref{sec:clinical}).
  \item \textbf{Concrete inclusive impact for named populations}, including a
  participant with a leg disability and a tracheostomy, and an estimate of the
  population served (Section~\ref{sec:impact}).
  \item \textbf{Operation suited to live performance}---nothing is worn on the
  performer's face, and setup is fast on an unmodified instrument---with
  public-concert evidence (Section~\ref{sec:eyesfree}).
  \item \textbf{A reproducible, open-source realization} with a real-time
  AR$\leftrightarrow$BLE rate-decoupling design and a non-destructive pneumatic
  anchor (Sections~\ref{sec:realization} and~\ref{sec:impl}).
\end{enumerate}

% =====================================================================
\section{Related work}
\label{sec:related}

\subsection{Assistive piano-pedal systems}
Several systems let players who cannot use their feet operate a piano pedal.
University of Victoria's CanAssist built a head-tilt ``Head-Activated Piano
Pedal'' for a wheelchair pianist: a floor device presses the pedal under control
of a headband sensor, engaging on a downward tilt and releasing on an upward
tilt~\cite{canassist}. Steingraeber \& S\"ohne offers, for wheelchair pianists,
electromagnetic pedal control integrated into the instrument, driven by helmet
inclination sensors, a neck bracelet with reed switches, or a bite-force sensor
that also conveys intermediate pedal positions~\cite{steingraeber,mouthpedal2008}.
Electromechanical pedal activators for players with disabilities appear in the
patent literature~\cite{hinsley1988,uspedal2008,uspedal2017}. bFaaaP differs by
(i)~using a \emph{commodity smartphone} as the head sensor; (ii)~a
\emph{quantitative, user-tunable} control law clinically tuned to the player;
(iii)~\emph{non-destructive} placement on an \emph{unmodified} acoustic piano via
pneumatic anchoring; and (iv)~being \emph{open source}.

\subsection{Robotic and mechatronic musical actuation}
Robotic and mechatronic systems perform music expressively, including
piano-playing robots and pedal automation~\cite{weinberg2020,scimeca2020,zappi2012,piano_manipulator2025,ctf_actuator2020}.
Such systems usually \emph{replace} the performer; bFaaaP instead \emph{augments}
a human performer, mapping the player's own head motion to the pedal so the
musical intention remains the player's.

\subsection{Hands-free, head-based human--computer interfaces}
Head movement (and gaze) is a well-studied accessible input for users with limited
hand use---for pointer control, software switches, and hands-free
operation~\cite{sumak2019,varona2008,manresa2014}---and commodity smartphone AR
frameworks now provide robust on-device face/head pose
estimation~\cite{arkit}. bFaaaP applies this idea to a continuous, expressive
musical control with performance-grade timing.

\subsection{Accessible digital musical instruments}
Accessible digital musical instruments (ADMIs) make music-making reachable for
people with motor disabilities through tangible, gaze, audio, and adapted
interfaces~\cite{frid2019,ramirez2023,duarte2023,eyeharp}. bFaaaP is complementary:
rather than a new instrument, it makes the \emph{conventional piano}'s pedal
accessible, on both acoustic and digital instruments.

% =====================================================================
\section{Results}
\label{sec:results}

\subsection{bFaaaP as a new inclusive design}
\label{sec:inclusive-design}
\begin{figure}[tbp]
\centering
\includegraphics[width=\linewidth]{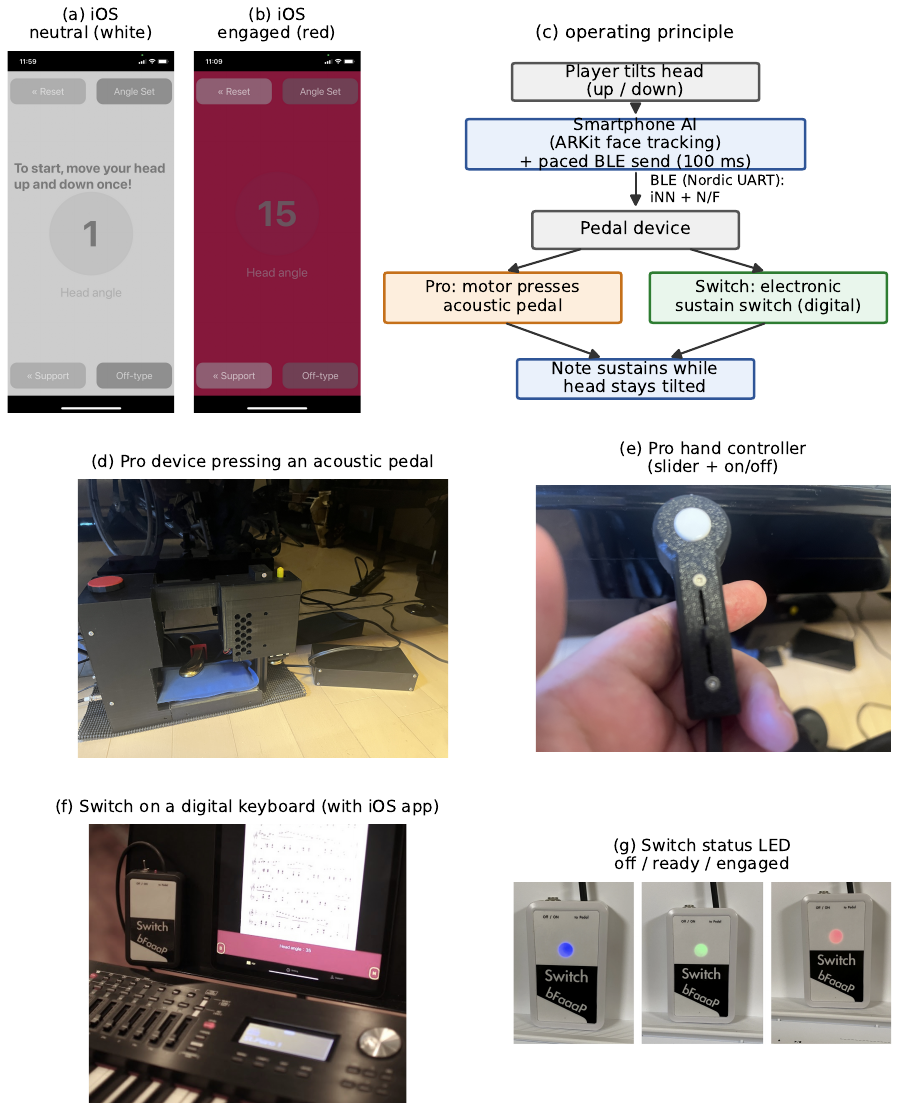}
\Description{Seven panels. (a) and (b) show the iOS app screen, white when neutral
and red when the pedal is engaged. (c) is a flow diagram: head tilt to smartphone
AI to BLE to the pedal device, branching to the Pro motor or the Switch electronic
output, producing a sustained note. (d) shows the Pro device pressing an acoustic
piano pedal on an anti-slip mat. (e) shows the hand controller with a slider and
button. (f) shows the Switch unit next to a digital keyboard with the app on a
tablet. (g) shows the Switch status LED in off, ready, and engaged colours.}
\caption{System overview. \textbf{(a,b)} The shared iOS controller's play screen
gives live feedback by colour: neutral (white/grey) below the head-angle threshold
(a) and red once the head is tilted past it so the pedal engages (b).
\textbf{(c)} Operating principle: a head tilt is sensed by smartphone AI (ARKit
face tracking) and relayed over a BLE link (Nordic UART Service) to the pedal
device, realized either as a motor that presses an acoustic pedal (\emph{Pro}) or
as an electronic sustain switch (\emph{Switch}). \textbf{(d)} The Pro device
pressing an acoustic piano's sustain pedal. \textbf{(e)} The Pro hand controller.
\textbf{(f)} The Switch unit beside a digital keyboard, with the same iOS app on
the tablet. \textbf{(g)} The Switch status LED mirrors the app state.}
\label{fig:overview}
\end{figure}

\begin{figure}[tbp]
\centering
\includegraphics[width=\linewidth]{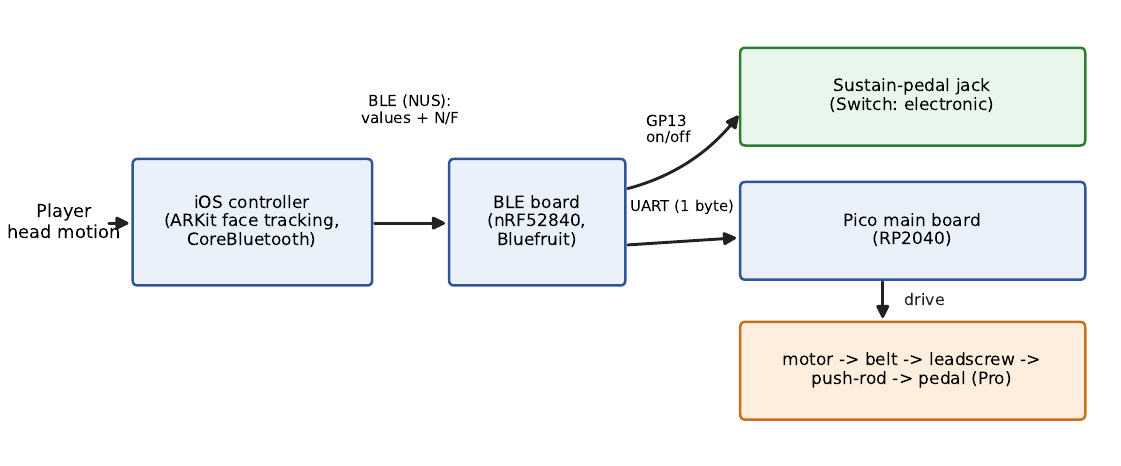}
\Description{Block diagram. Player head motion feeds the iOS controller (ARKit face
tracking, CoreBluetooth), which sends BLE Nordic UART values plus engage/release
signals to an nRF52840 BLE board. The BLE board either drives the sustain-pedal
jack directly for the Switch line, or forwards one byte over UART to a Pico RP2040
main board that drives the motor, belt, lead screw, and push-rod for the Pro line.}
\caption{System architecture. One smartphone controller drives two co-equal device
realizations (Switch: electronic; Pro: robotic).}
\label{fig:arch}
\end{figure}

bFaaaP is one smartphone controller plus two co-equal device realizations
(Figures~\ref{fig:overview} and~\ref{fig:arch}). The smartphone
(iOS; an iPhone/iPad with a TrueDepth front camera) tracks the player's head
pitch with AR face tracking and computes a control value, which it transmits over
a Nordic UART Service (NUS) BLE link. The device then operates the pedal so a note
sustains while the head is tilted past a threshold and releases when the head
returns. The same controller and control law serve \textbf{Pro} (a motor presses
an \emph{acoustic} piano's sustain pedal) and \textbf{Switch} (the sustain of a
\emph{digital} instrument is actuated electronically through its pedal jack).
We treat the two as co-equal: Pro restores pedalling on the
instruments where it is hardest (unmodified grands and uprights), while Switch's
simplicity gives it very broad, low-cost reach across digital pianos and
keyboards. The design intent throughout is to reproduce the player's
\emph{intended} pedalling, not merely to enable an on/off effect.

\subsection{The contribution is the control law, not the architecture}
\label{sec:control-law}
\begin{figure}[tbp]
\centering
\includegraphics[width=\linewidth]{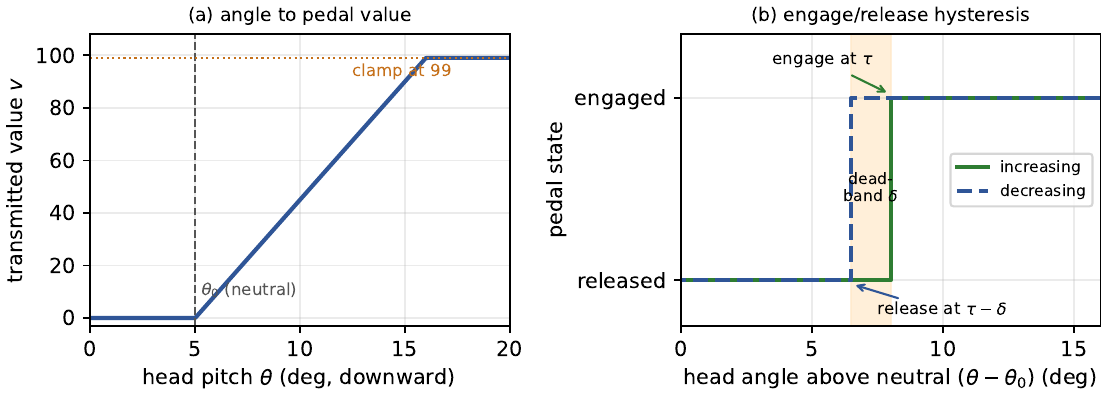}
\Description{Two plots. Left: transmitted value versus head angle above the offset,
rising linearly with the multiplier and clamped at 99. Right: pedal state versus
head angle showing a hysteresis loop with a dead-band between the engage and
release thresholds.}
\caption{Head-angle control law (Eq.~\ref{eq:law}). (a) The transmitted value maps
the head angle above the offset, scaled by the user multiplier and clamped.
(b) Engage/release use hysteresis with a small dead-band to avoid chatter.}
\label{fig:control}
\end{figure}

Let $\theta$ be the (downward) head pitch reported by the face tracker and
$\theta_{\mathrm{off}}$ a user-calibrated offset. The transmitted pedal value is
\begin{equation}
v = \mathrm{clamp}\big(\,(\theta-\theta_{\mathrm{off}})\cdot m,\;0,\;99\,\big),
\label{eq:law}
\end{equation}
where $m$ is a user-selected multiplier. The \emph{recommended} parameter ranges,
validated in our study, are a user offset of $3$--$10^\circ$ and a multiplier of
$10$--$50$, chosen so that a further $2$--$10^\circ$ of tilt reaches full pedal
action (Figure~\ref{fig:control}a); as Table~\ref{tab:participants} reports,
participants fine-tuned within a slightly wider observed span (offset
$3$--$19^\circ$, multiplier $8$--$50$). Engage/release events use \emph{hysteresis}: engage
when $\theta\ge\theta_{\mathrm{off}}$ while released, release when
$\theta<\theta_{\mathrm{off}}-\delta$ while engaged, with a small dead-band
$\delta$ (Figure~\ref{fig:control}b). The player presets just two quantities to
taste: the \emph{offset} (\emph{where} the response begins) and the
\emph{multiplier} (\emph{how much} pedal value each degree of tilt produces). These
two presets in turn fix a \emph{secondary, temporal} characteristic we call the
\emph{response speed}---\emph{how fast} the pedal actuator follows the head once it
is past the offset. A small offset with a large multiplier, for example, gives a
fast response that suits a player with a restricted range of head motion (on the Pro
device this also raises the actuator's commanded movement speed). The player thus
realizes a preferred response speed \emph{indirectly}, by pre-setting the offset and
multiplier rather than by setting a separate ``speed'' control.

\begin{figure}[tbp]
\centering
\includegraphics[width=0.92\linewidth]{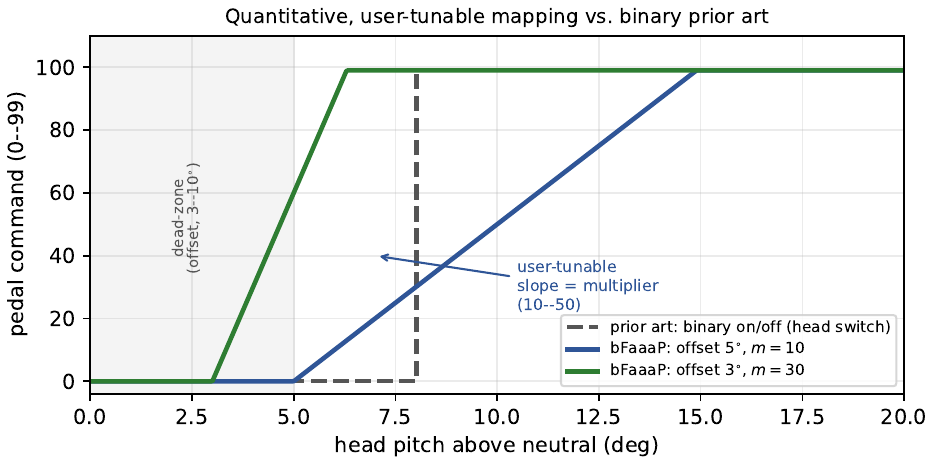}
\Description{Line plot of pedal command versus head pitch above neutral. A dashed
grey step shows binary on/off prior art switching fully on at a single threshold.
Two solid lines show bFaaaP with a dead-zone offset and a user-tunable slope
(multiplier): one with offset five degrees and multiplier ten, one with offset
three degrees and multiplier thirty, both rising proportionally and clamping at the
maximum.}
\caption{The novelty is quantitative and user-tunable. Binary prior art emits a
fixed on/off at one threshold; bFaaaP yields a proportional command whose dead-zone
(offset $3$--$10^\circ$) and slope (multiplier $10$--$50$) the user sets.}
\label{fig:cvp}
\end{figure}

This quantitative, user-tunable mapping---not the head$\rightarrow$pedal
architecture---is the inventive core. Patent examination of the underlying method
found the bare architecture (a head sensor, a processor, an actuator, and a pedal)
anticipated or obvious in view of a head-activated piano pedal~\cite{canassist}
combined with head-gesture display, head-motion, and headset prior
art~\cite{hmd_jp2017,headset_jp2012,colopl_jp2017,titech_jp2014,alps_us2018}, and
with an existing pedal-performance aid~\cite{yamaha_jp2006};
the claims were granted on the \emph{specific} offset/multiplier law and the
pre-adjustable response speed it yields~\cite{jp6726319,jp7004771}. Figure~\ref{fig:cvp}
contrasts the two: where binary prior art emits a fixed on/off at a single
threshold, bFaaaP produces a continuous, proportional command whose dead-zone and
slope each user pre-sets to taste. This per-player pre-setting of the offset and
multiplier yields an \emph{individualized control}---which is precisely what lets
very different players (Section~\ref{sec:impact}) reproduce their intended
pedalling. The mapping is realized in software, and the per-player preset is a
one-time deployment step, as described in Sections~\ref{sec:law-sw}
and~\ref{sec:impl-preset}.

\subsection{Clinical validation: the Auxiliary Pedal Effect Evaluation (APEE)}
\label{sec:clinical}

\begin{figure}[tbp]
\centering
\includegraphics[width=\linewidth]{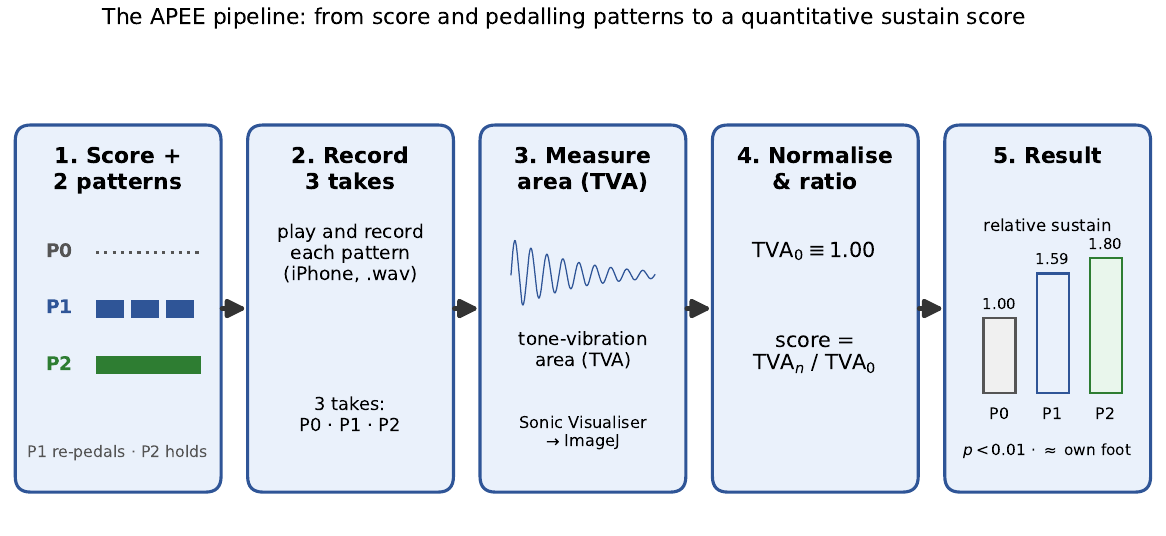}
\Description{A five-stage pipeline. Stage 1: the test motif with pedalling patterns
0 (no pedal), 1 (re-pedal at each three-note group), and 2 (held across groups).
Stage 2: each pattern is played and recorded. Stage 3: the tone-vibration area (TVA)
of each recording's waveform is measured with Sonic Visualiser and ImageJ. Stage 4:
the no-pedal area is set to 1.00 and the score is the ratio TVA_n over TVA_0.
Stage 5: the result, relative sustain 1.00, 1.59, 1.80, significant at p below 0.01
and not significantly different from the player's own foot.}
\caption{The APEE method end to end: from the test score and its two pedalling
patterns, through recording and tone-vibration-area (TVA) measurement, to the
normalized sustain score (TVA$_n$/TVA$_0$) and the result.}
\label{fig:apee_pipeline}
\end{figure}

\begin{figure}[tbp]
\centering
\includegraphics[width=\linewidth,trim=0 0 0 70,clip]{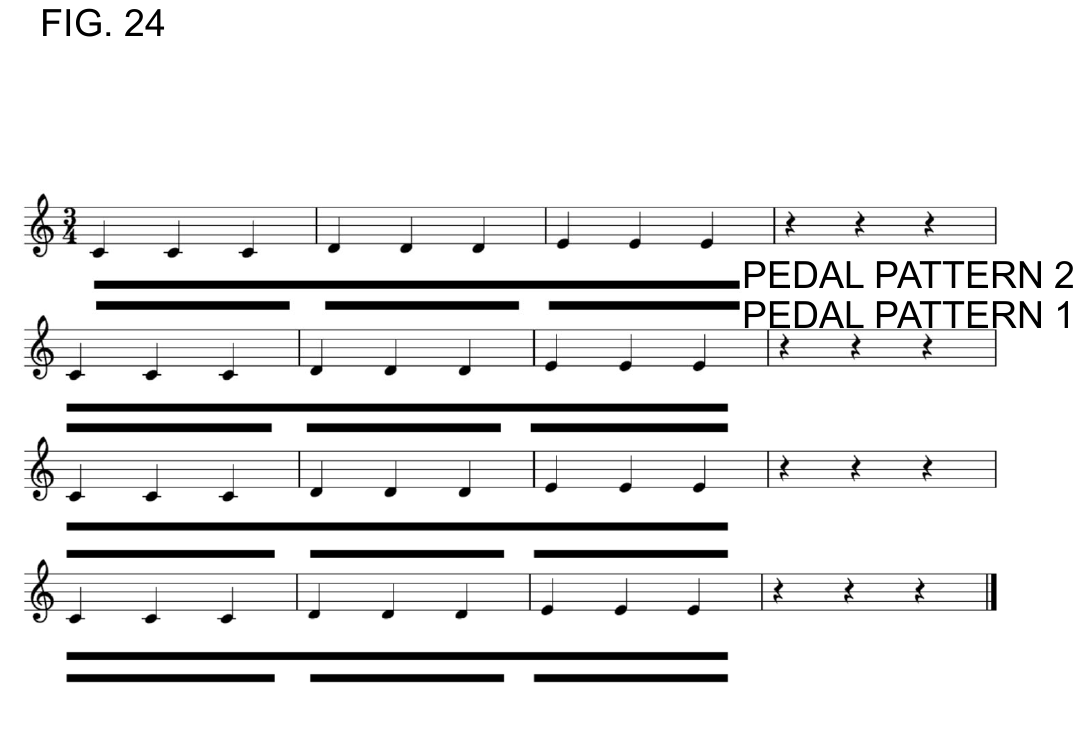}
\Description{The study score: the motif do-do-do, re-re-re, mi-mi-mi in 3/4 time,
repeated four times. Under each system, two rows of horizontal pedal-down spans mark
the two pedalling patterns: the lower row (pattern 1) is broken into one span per
three-note group, i.e. the pedal is re-pressed at each group; the upper row
(pattern 2) is a single long span held across the groups.}
\caption{The APEE test score and its two pedalling patterns (the actual study score,
reproduced from the patent application, PCT Fig.~23/24). The motif
(do--do--do, re--re--re, mi--mi--mi; $3/4$, repeated four times) is played three ways.
The horizontal bars beneath each staff are the pedal-down spans: \textbf{pattern~1}
(lower row) re-pedals at each three-note group, whereas \textbf{pattern~2} (upper row,
one long bar) holds the pedal across the groups for a longer, more connected
sustain---which is why pattern~2 yields more sustained-tone energy than pattern~1
(Table~\ref{tab:sustain}). \emph{Pattern~0} (not drawn) is the same motif played with
no pedal.}
\label{fig:apee_score_real}
\end{figure}

\begin{figure}[tbp]
\centering
\includegraphics[width=\linewidth]{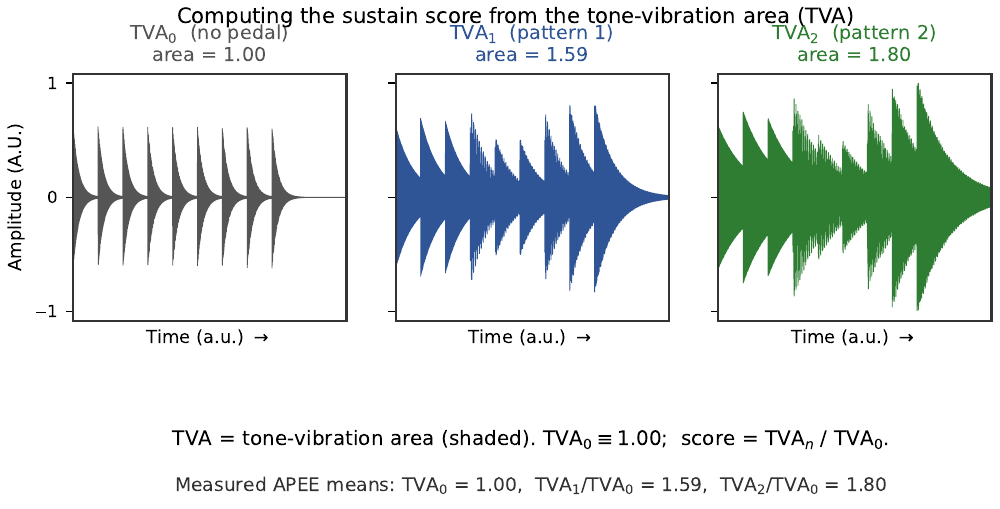}
\Description{Three waveform panels for the same motif: no pedal (TVA0), bFaaaP
pattern 1 (TVA1), and pattern 2 (TVA2). The shaded, rectified waveform area grows
with sustain. With TVA0 set to 1.00, each pattern's score is TVA_n over TVA_0;
the measured means are 1.00, 1.59, and 1.80.}
\caption{Computing the sustain score from the tone-vibration area (TVA). Each
recording's waveform area (shaded) is measured; with the no-pedal area set to
$1.00$, each pattern's score is TVA$_n$/TVA$_0$. Measured means: $1.00$, $1.59$,
$1.80$.}
\label{fig:tva}
\end{figure}

The control law and its parameter ranges were established and validated in a
human-subject study, the Auxiliary Pedal Effect Evaluation (APEE), reported in the
patent specification~\cite{jp6726319} and summarized here;
Figure~\ref{fig:apee_pipeline} shows the method end to end. Fifteen participants
were grouped into three classes (Table~\ref{tab:participants}): Class~I, adults who
\emph{can} use their own feet (controls); Class~II, children whose feet do not
reach the pedals; and Class~III, people with leg/neck disabilities. Each
participant practised a fixed motif and recorded it under three conditions:
\emph{pattern~0}, the motif with no sustain pedal (the baseline); and
\emph{pattern~1} and \emph{pattern~2}, the \emph{same} motif played with bFaaaP
under two prescribed sustain-pedalling patterns marked in the study score
(Figure~\ref{fig:apee_score_real}): pattern~1 re-pedals at each three-note group,
whereas pattern~2 holds the pedal across the groups for a longer, more connected
sustain, so pattern~2 is the more sustained of the two. Class~I additionally
recorded with their
own foot as a control. Each recording was analysed by measuring a tone-vibration
area (TVA) from the audio (via Sonic Visualiser and ImageJ), with pattern~0
normalized to $1.00$ (Figure~\ref{fig:tva}).

\begin{table}[tbp]
\caption{APEE participants ($n=15$), grouped by class, with the offset and
multiplier ranges they chose. Overall, offsets spanned $3$--$19^\circ$ (with
$5$--$10^\circ$ preferred) and multipliers $8$--$50$ ($20$--$40$ preferred).}
\label{tab:participants}
\centering
\small
\begin{tabular}{@{}llp{4.6cm}cc@{}}
\toprule
Class & $n$ & Participants & Offset (deg) & Multiplier \\
\midrule
I (control) & 7 & adults able to use their own feet (incl.\ one elderly) & 5--19 & 10--40 \\
II & 5 & children (ages 7--11) & 5 & 20--30 \\
III & 3 & adults with disabilities: leg disability ($\times2$); leg disability + tracheostomy ($\times1$) & 3--10 & 8--50 \\
\bottomrule
\end{tabular}
\end{table}

\begin{figure}[tbp]
\centering
\includegraphics[width=\linewidth]{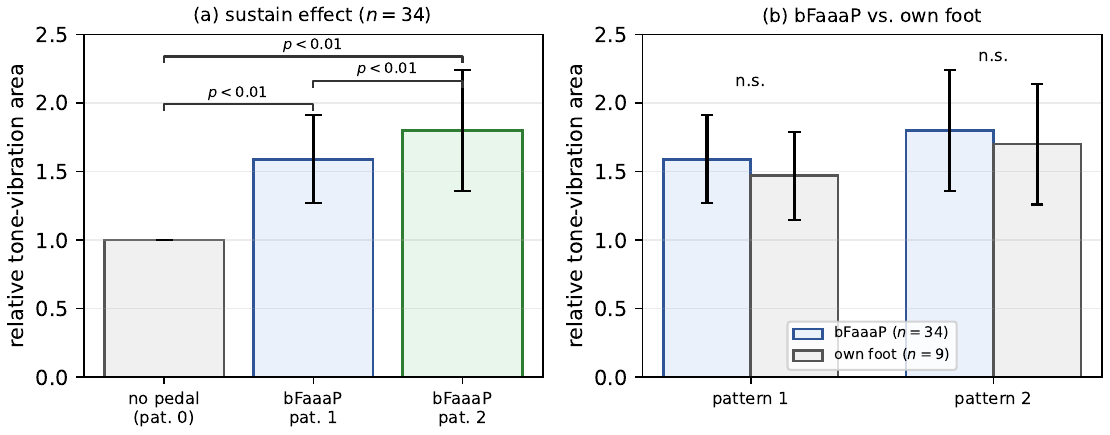}
\Description{Two bar charts. (a) Sustain effect: relative tone-vibration area for
no pedal (1.00), bFaaaP pattern 1 (1.59), and bFaaaP pattern 2 (1.80), with error
bars; all pairwise comparisons marked p less than 0.01. (b) bFaaaP versus own foot
for patterns 1 and 2, with overlapping error bars marked not significant.}
\caption{APEE clinical results. (a) Both bFaaaP patterns significantly increase the
sustained-tone energy over no pedal ($p<0.01$). (b) bFaaaP is statistically
indistinguishable from the player's own foot (n.s.).}
\label{fig:apee}
\end{figure}

\paragraph{bFaaaP produces a strong, controllable sustain.}
Relative to no pedal, both bFaaaP patterns significantly increased the
sustained-tone energy, and the two patterns differed from each other
(Table~\ref{tab:sustain}, Figure~\ref{fig:apee}a)---evidence that the device gives
not just ``a'' sustain but a \emph{controllable} one.

\begin{table}[tbp]
\caption{Sustain effect: tone-vibration area relative to no pedal ($n=34$ bFaaaP
recordings). All paired comparisons are significant ($p<0.01$).}
\label{tab:sustain}
\centering
\small
\begin{tabular}{@{}lcc@{}}
\toprule
Condition & Mean (rel.\ TVA) & SD \\
\midrule
Pattern 0 (no pedal)   & 1.00 & 0.00 \\
Pattern 1 (bFaaaP)     & 1.59 & 0.32 \\
Pattern 2 (bFaaaP)     & 1.80 & 0.44 \\
\midrule
\multicolumn{3}{@{}l}{Paired $t$-test: P0--P1 $p=4.6{\times}10^{-12}$; P0--P2 $p=7.3{\times}10^{-12}$;}\\
\multicolumn{3}{@{}l}{P1--P2 $p=2.3{\times}10^{-4}$.}\\
\bottomrule
\end{tabular}
\end{table}

\paragraph{bFaaaP is statistically equivalent to the player's own foot.}
For Class~I participants, who could pedal with their own foot, the sustain
achieved with bFaaaP did not differ significantly from the sustain achieved with
their foot (Table~\ref{tab:vsfoot}, Figure~\ref{fig:apee}b). In other words,
bFaaaP reproduces foot-quality pedalling---the basis for our claim that it
realizes the player's \emph{intended} pedalling rather than a coarse substitute.

\begin{table}[tbp]
\caption{bFaaaP vs.\ the player's own foot (Class~I). No significant difference
(two-sided $t$-test; equal variances not rejected, $F$-test $p>0.05$).}
\label{tab:vsfoot}
\centering
\small
\begin{tabular}{@{}lccc@{}}
\toprule
Condition & bFaaaP ($n{=}34$) & own foot ($n{=}9$) & $t$-test $p$ \\
\midrule
Pattern 1 & 1.59 & 1.47 & 0.33 (n.s.) \\
Pattern 2 & 1.80 & 1.70 & 0.43 (n.s.) \\
\bottomrule
\end{tabular}
\end{table}

\paragraph{Piano experience did not matter; instrument did.}
Splitting participants by piano experience ($\ge5$ vs.\ $<5$ years) showed no
significant score difference ($t$-test $p>0.05$), indicating the device is usable
without extensive training; differences \emph{between pianos} were significant
(one-way ANOVA, $p<0.05$), as expected for different instruments.

\subsection{Concrete inclusive impact across populations}
\label{sec:impact}
A central accessibility finding is that the sustain effect did \emph{not} differ
significantly across the three participant classes (Table~\ref{tab:classes}):
adults, children, and people with disabilities all benefited comparably. bFaaaP is
thus not a narrow aid but a broadly inclusive one.

\begin{table}[tbp]
\caption{Sustain effect across participant classes (one-way ANOVA). No significant
difference between adults (I), children (II), and people with disabilities (III).}
\label{tab:classes}
\centering
\small
\begin{tabular}{@{}lcccc@{}}
\toprule
Pattern & Class I & Class II & Class III & ANOVA $p$ \\
\midrule
Pattern 1 & 1.54 ($n{=}18$) & 1.78 ($n{=}10$) & 1.43 ($n{=}6$) & 0.068 (n.s.) \\
Pattern 2 & 1.75 ($n{=}18$) & 1.91 ($n{=}10$) & 1.75 ($n{=}6$) & 0.66 (n.s.) \\
\bottomrule
\end{tabular}
\end{table}

The user-tunable control law is what makes this breadth possible. One Class~III
participant had a leg disability \emph{and} a tracheostomy, which restricted her
range of head motion; she chose a \emph{small} offset ($3$--$5^\circ$) and a
\emph{large} multiplier ($40$--$50$) so that her limited tilt still spanned the
full pedal, and performed the target passage successfully. The population that
benefits is large and worldwide. Long-term ventilation---often delivered through a
tracheostomy, which restricts head motion---reaches, for example, about $7{,}700$
people on home tracheostomy positive-pressure ventilation in Japan as of 2020 (and
$\sim$21{,}000 on home ventilation overall)~\cite{hmv_japan2020}, around $21{,}500$
home-ventilation users across the 16 European countries of the Eurovent
survey~\cite{eurovent2005}, and $4{,}334$ in Canada~\cite{hmv_canada2015}; reported
home-ventilation prevalence runs about $6.6$--$13$ per $100{,}000$ in high-income
countries and has been rising. More broadly still, the people who cannot readily use
a foot pedal are far more numerous: the World Health Organization estimates that
about $80$~million people---roughly $1\%$ of the world's population---need a
wheelchair~\cite{who_wheelchair2023}. A design that explicitly serves restricted head
motion, and that runs on a commodity smartphone, therefore has real, global reach.

A short post-study questionnaire (Table~\ref{tab:survey}) found most participants
enjoyed bFaaaP and rated it positively; the most common reservation concerned
ease of use, consistent with participant comments that releasing the pedal is
slightly harder than pressing it and that ``the more you practise, the easier it
becomes.''

\begin{table}[tbp]
\caption{Post-study questionnaire ($n=15$; counts of respondents).}
\label{tab:survey}
\centering
\small
\begin{tabular}{@{}lccc@{}}
\toprule
Question & Positive & Neutral & Negative \\
\midrule
Was it enjoyable?     & 9.0 & 5.0 & 1 \\
Easy to use?          & 3.0 & 10.0 & 2 \\
Does it look good?    & 4.5 & 7.5 & 3 \\
Is the name good?     & 8.0 & 6.0 & 1 \\
\bottomrule
\end{tabular}
\end{table}

\subsection{Practical fit for live performance: nothing worn on the face, fast setup}
\label{sec:eyesfree}
A design insight, sharpened during patent examination, is that head-gesture
interfaces that depend on a \emph{visual indicator}---for example, a head-mounted
display that shows the wearer whether a threshold has been crossed~\cite{hmd_jp2017}---are
ill-suited to a \emph{performing pianist}, who must keep looking at the score, the
keys, and the audience and cannot wear a display while playing. bFaaaP places the
sensor off the body entirely: the smartphone rests on the \emph{music stand} and
reads the player's head angle from there, so \emph{nothing is attached to the
performer's face or head}. This is a deliberate contrast with display- or
headband-based prior art---a head-mounted display, or CanAssist's head-tilt pedal
for the wheelchair pianist ``Emily,'' which sensed head motion with a sensor worn on
a \emph{headband}~\cite{canassist}. The player simply feels the sustain in the sound;
the only optional cue (the app screen turning red, or the Switch LED) is a visual aid
for a helper or for setup, not something the performer must watch.

For live use, the \emph{ease and speed of setup} matter as much as the interaction
itself, because a concert stage is shared and the changeover between performers is
short. Because bFaaaP is simply placed---the Switch lead plugged into the digital
instrument, or the Pro unit rested over the pedals and anchored by the airback on an
\emph{unmodified} acoustic piano---with nothing worn by the player, it can be installed
quickly, within the same changeover window as the other performers. We have validated
this in the field rather than only in the lab: bFaaaP has appeared in piano recitals,
university festivals, an international assistive-equipment exhibition, and full public
concerts from 2018 onward, with players who cannot use the foot pedal---including a
wheelchair user---performing pieces that require sustain before live
audiences~\cite{repo,videos}. A representative \emph{formal} multi-performer example
is the seven-university joint concert \emph{Natsu-no-Kyoen 2022} (Toyosu Civic
Center Hall, Tokyo, 5~September 2022), where a Tokyo~Tech Platanus member
performed the piano piece \emph{Blumenlied} (``Flower Song'') with bFaaaP, set up
among the other performers on the shared stage~\cite{concert_kyoen2022}---the same
tight changeover faced by every performer. The on-instrument setup itself is
documented in the 2025 Suzukake Science-Day concert recording (the first author's
walkthrough at 25:01--28:32)~\cite{concert_suzukake2025}, with a dedicated,
step-by-step Pro installation guide also available~\cite{video_pro_setup}.

% --- 3.6 System realization (drafted to sub-subsection level) ---
% =====================================================================
% Draft section for 作業7 (paper full revision) — Results §3.6
% "System realization" — drafted to sub-subsection level with content.
% LaTeX is written to compile in BOTH the arXiv (article) and the
% TACCESS (acmart) manuscripts. Cross-refs use existing labels:
%   fig:overview, fig:arch, fig:control, fig:rate, fig:pro
% Citations use existing keys: arkit, nus, rp2040, repo.
% Degrees are written as ^{\circ} (math) and multiply as \times so no
% extra package is required.
% =====================================================================

\subsection{System realization}
\label{sec:realization}

The clinically validated control law of Section~\ref{sec:clinical} is realized
end to end by \emph{one} smartphone controller and \emph{two} interchangeable
device families (Pro for acoustic pianos, Switch for digital instruments), joined
by a deliberately simple wireless protocol and, on the Pro device, a short wired
link that separates two timing domains. This subsection describes each stage and
where the quantitative parameters (user offset $3$--$10^{\circ}$,
multiplier $10$--$50$, full action within $+2$--$10^{\circ}$, and the response
speed that those two presets produce) physically live in the stack.

\subsubsection{Realized pipeline: one controller, two devices, a wired bridge}
\label{sec:pipeline}
Figure~\ref{fig:arch} summarizes the realized data path. The iOS app extracts a
head-tilt angle, applies the control law, and transmits compact messages over
Bluetooth Low Energy (BLE). On the \emph{Switch} device a single
microcontroller turns those messages directly into an electronic sustain
on/off; on the \emph{Pro} device a BLE board relays a one-byte command over a
wired UART to a second microcontroller that drives a motorized actuator against
the physical pedal. The same controller, BLE protocol, and control law serve
both devices, so an implementer learns one interaction model
(Figure~\ref{fig:overview}).

\subsubsection{iOS controller: AR face tracking and head-tilt extraction}
\label{sec:ios-ar}
The controller uses ARKit face tracking on a TrueDepth (Face~ID-class) device to
estimate head pose at roughly $60$\,fps~\cite{arkit}. A session delegate updates
a single shared head-tilt variable on every frame; the downward pitch component
is the control signal. Because face tracking is hardware-bound, the controller
runs on a physical iPhone/iPad rather than a simulator, and the screen is kept
awake during play so the foot-free session is uninterrupted.

\subsubsection{The quantitative control law in software}
\label{sec:law-sw}
The controller implements the quantitative mapping of
Section~\ref{sec:control-law} in software: each transmitted value is
$\mathrm{clamp}((\theta-\theta_{\mathrm{off}})\cdot m,\,0,\,99)$ with the user's
offset $\theta_{\mathrm{off}}$ (validated $3$--$10^{\circ}$) and multiplier $m$
(validated $10$--$50$), so a further $2$--$10^{\circ}$ of tilt reaches full pedal
action (Figure~\ref{fig:control}). Engagement uses
hysteresis: an \emph{engage} event is emitted when $\theta\ge\theta_{\mathrm{off}}$
while the pedal is released, and a \emph{release} event when $\theta$ falls below
$\theta_{\mathrm{off}}$ by a small dead-band while the pedal is engaged. Because
the events are edge-triggered with a state guard, one head dip produces exactly
one engage and one release, eliminating chatter near the threshold. The
\emph{response speed} (how fast the command tracks the head beyond the offset) is
not a separate control but a secondary, temporal effect of the user's \emph{offset}
and \emph{multiplier} presets---which is what lets users with a restricted range of
motion (small offset, large multiplier) obtain a fast response.

\subsubsection{Decoupling AR sampling from BLE transmission}
\label{sec:decoupling}
\begin{figure}[tbp]
\centering
\includegraphics[width=\linewidth]{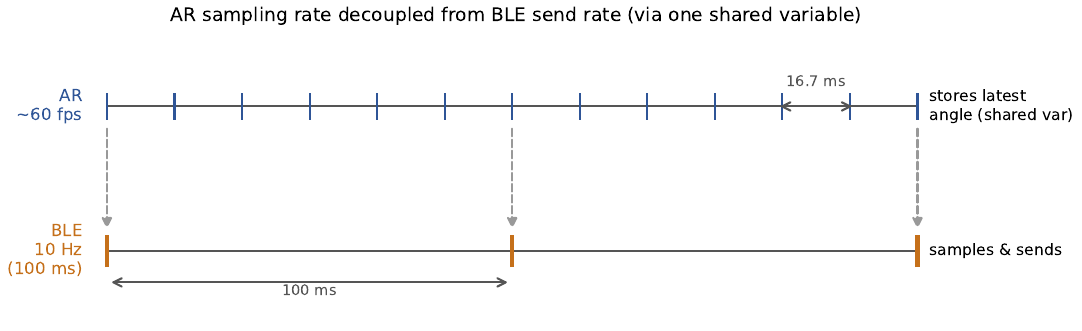}
\Description{Timeline comparing a dense series of AR frames at about 60 frames per
second with a sparse series of BLE transmissions at 10 hertz; the tracker updates a
shared variable that the slower timer samples and transmits.}
\caption{The fast AR sampling rate ($\sim$60\,fps) is decoupled from the slower BLE
send rate ($\sim$10\,Hz): the tracker only updates a shared variable, which a timer
samples and transmits.}
\label{fig:rate}
\end{figure}

The face tracker produces samples far faster than a BLE link should be driven, so
the controller \emph{decouples the two rates} (Figure~\ref{fig:rate}). The
continuous value is not sent per frame; a fixed-period timer ($\approx$100\,ms,
i.e.\ $10$\,Hz) reads the latest shared head-tilt variable and transmits it, while
the engage/release events are emitted on crossing with a short ($\approx$10\,ms)
minimum spacing. Historically the transmit period was tuned upward (from a few
milliseconds to $\approx$100\,ms) to remove a congestion/ordering fault on the
link. This producer/consumer decoupling between a fast AI sampler and a
slower wireless transport is the system's central real-time-engineering
contribution and is reused on the device side (Section~\ref{sec:coordination}).

\subsubsection{BLE link and the wire protocol}
\label{sec:ble}
The controller is a BLE central that scans, \emph{name-filters} the known device
names (\texttt{bFaaaPSwitch\_1}\allowbreak..\allowbreak\texttt{\_4}) so it never attaches to unrelated
peripherals, connects to the user-selected channel, and writes over the Nordic
UART Service (NUS)~\cite{nus} using write-without-response for low latency. The
protocol is intentionally tiny: a continuous value (\texttt{i00}..\texttt{i99}),
engage/release events, channel-rename commands, and an on-type/off-type toggle.
The play screen turns from white to red once the head passes the threshold,
giving a visual confirmation (for a helper or for setup) that matches the
device-side indicator (Section~\ref{sec:switch}).

\subsubsection{Pro device: BLE board $\rightarrow$ microcontroller $\rightarrow$ drivetrain $\rightarrow$ pedal}
\label{sec:pro-chain}
\begin{figure}[tbp]
\centering
\includegraphics[width=0.92\linewidth]{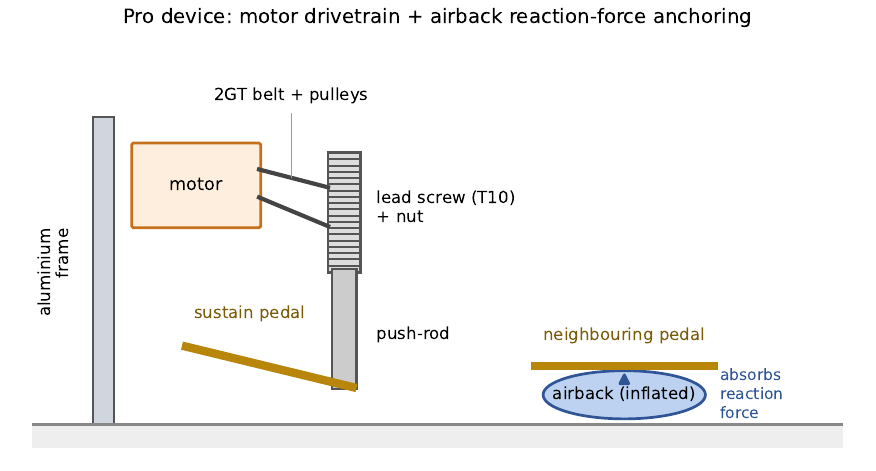}
\Description{Side-view schematic of the Pro device. A motor on an aluminium frame
drives a timing belt and a lead screw, which lowers a push-rod onto the sustain
pedal. An inflated airback under a neighbouring pedal absorbs the upward reaction
force and anchors the unit.}
\caption{Pro device. A motor drives a timing belt and lead screw that advances a
push-rod onto the sustain pedal; an inflatable airback beneath a neighbouring pedal
absorbs the reaction force and anchors the unit on an unmodified piano.}
\label{fig:pro}
\end{figure}

The Pro device uses two microcontrollers. An nRF52-class BLE board receives the
NUS messages and forwards a one-byte command over a wired UART to an RP2040 main
board~\cite{rp2040}. The RP2040 maps the byte to a target position between
calibrated travel limits, with a small deadband so sub-threshold jitter does not
move the motor, and drives a motor through a timing belt and a vertical lead screw
that advances a push-rod straight down against the sustain pedal (Figure~\ref{fig:pro}). The
result is continuous, proportional pedal depth that follows the head angle.

\subsubsection{Reaction-force anchoring (airback) and sensorless self-calibration}
\label{sec:airback}
Two device-side mechanisms make the Pro practical on an \emph{unmodified}
acoustic piano. First, an \emph{airback} (a term we coined for an inflatable,
air-braced anchor---\emph{not} a safety ``airbag''): at power-up a pump (driven through a
power MOSFET) inflates an air bag (a commercial air-wedge cushion) placed under an
adjacent pedal until a pressure threshold is reached, absorbing the actuator's reaction force so the lightweight
device stays put without being bolted to the instrument (a non-destructive
anchoring). Second, \emph{sensorless self-calibration}: the actuator advances
until the motor's electrical \emph{power} --- proportional to the reaction force,
and chosen over current because power is robust to supply-voltage variation
(current alone changes with the input voltage) --- reaches a limit, which
automatically fixes the lower travel end; the upper end is set live by a slider on
a small hand controller. These two ends correspond directly to the offset (zero)
and the full-press depth of the control law.

\subsubsection{Switch device: electronic sustain switching for digital instruments}
\label{sec:switch}
For digital pianos and keyboards the Switch device needs no motor, drivetrain, or
airback. The BLE board drives a GPIO pin into a low-side MOSFET that opens or
closes the instrument's sustain jack, so the same engage/release events that move
a motor on the Pro instead toggle an electronic sustain. An on-board RGB LED
mirrors the controller's white$\rightarrow$red feedback (idle / armed / engaged),
and the on-type/off-type toggle accommodates the jack polarity of different
instruments.

\subsubsection{End-to-end coordination and timing budget}
\label{sec:coordination}
On the Pro device the wired UART bridges an \emph{asynchronous} wireless domain
(the BLE board, whose radio stack is event-driven) and a \emph{synchronous}
motion domain (the RP2040, which runs a deterministic blocking loop) --- the same
decoupling philosophy as the iOS AR/BLE split (Section~\ref{sec:decoupling}). The
budget holds in the field: in a real grand-piano session the controller emitted
on the order of two thousand continuous values and an equal count of engage and
release events (one per head dip) with no transmission errors and no
desynchronization, confirming that the $10$\,Hz value stream plus throttled
events stay within what the link and the actuator can absorb.

% =====================================================================
\section{Discussion}
\label{sec:discussion}

\subsection{Inclusive design and human--machine interaction}
bFaaaP reframes pedal access from ``enable on/off'' to ``reproduce intended
pedalling.'' The clinical equivalence to the player's own foot
(Table~\ref{tab:vsfoot}) and the controllable difference between pedalling
patterns (Table~\ref{tab:sustain}) indicate that the head channel carries genuine
musical intention, not just a trigger. As a human--machine interaction this is a
favourable trade: a single, natural degree of freedom (head tilt), made expressive
by a quantitative law the user owns.

\subsection{Accessibility impact across populations}
Because the effect is statistically uniform across adults, children, and people
with disabilities (Table~\ref{tab:classes}), bFaaaP is inclusive by evidence, not
just by aspiration. The tunable law extends the reach to severe cases such as
restricted head motion from a tracheostomy (Section~\ref{sec:impact}), and the
two device lines cover both the acoustic instruments where pedalling is hardest to
restore and the large installed base of digital pianos.

\subsection{Population scale, and the controller as an accessibility-control channel}
\label{sec:access-channel}
The people whom a foot-free, head-controlled interface could serve are numerous and
worldwide, and two largely distinct populations are illustrative. First, people who
cannot readily use a foot pedal because of lower-limb impairment overlap strongly with
\emph{wheelchair users}: the World Health Organization estimates that $\sim$80~million
people---about $1\%$ of the world's population---need a wheelchair~\cite{who_wheelchair2023},
and national surveys report from hundreds of thousands to millions of users in individual
countries (Table~\ref{tab:wheelchair}). Second, people with \emph{restricted head
motion}---for example from a tracheostomy and long-term ventilation, the hardest case our
study addressed---number in the thousands to tens of thousands per country on home
ventilation alone (Table~\ref{tab:hmv}). We caution that the figures in both tables come
from heterogeneous surveys with different definitions and metrics and are \emph{not}
strictly comparable; we report them to convey scale, not for precise cross-country
ranking, and we note that WHO publishes only a single \emph{global} wheelchair estimate
rather than a country-by-country table~\cite{who_great2022}.

\begin{table}[tbp]
\caption{Wheelchair users (or people who \emph{need} a wheelchair), by region. Metrics,
definitions, and years differ and are not strictly comparable (see note).}
\label{tab:wheelchair}
\centering
\small
\setlength{\tabcolsep}{4pt}
\begin{tabular}{@{}lp{5.2cm}ll@{}}
\toprule
Region & Estimate & Year & Source \\
\midrule
World & $\sim$80 million ($\sim$1\% of population) \emph{need} a wheelchair & --- & WHO~\cite{who_wheelchair2023,who_great2022} \\
USA & 3.6 million users (1.5\%, aged 15+) & 2010 & US Census~\cite{us_census_disability2010} \\
UK (England) & $\sim$1.2 million users (estimate) & 2017 & NHS England~\cite{nhs_wheelchair} \\
Canada & 288{,}800 wheelchair/scooter users ($\sim$1.0\%) & 2012 & \cite{wc_canada_smith2016} \\
Japan & $\sim$818{,}000 manual wheelchairs in use ($\sim$0.6\%) & 2019 & \cite{wc_japan_shirogane2019} \\
Australia & $\sim$119{,}000 manual-wheelchair users (65+); 679{,}000 use mobility aids & 2018 & AIHW~\cite{wc_australia_aihw} \\
\bottomrule
\end{tabular}

\smallskip
{\footnotesize \emph{Note.} Definitions differ (``users'' vs.\ ``need''; community-only
vs.\ all; age cut-offs), so totals are not strictly comparable. Germany has no official
wheelchair-user count. WHO gives only a single global estimate.}
\end{table}

\begin{table}[tbp]
\caption{Home mechanical ventilation (HMV) and its invasive/tracheostomy subset, by
country. Metric \emph{types} differ (see note).}
\label{tab:hmv}
\centering
\small
\setlength{\tabcolsep}{4pt}
\begin{tabular}{@{}lrll l@{}}
\toprule
Country & HMV users & Invasive (trach.) & /100k & Source \\
\midrule
Japan & $\sim$21{,}000 & 7{,}700 (TPPV) & --- & 2020~\cite{hmv_japan2020} \\
Europe (16 countries) & 21{,}526 & varies & 6.6 & 2002~\cite{eurovent2005} \\
Canada & 4{,}334 & $\sim$18\% & 12.9 & 2012~\cite{hmv_canada2015} \\
Poland & 12{,}616 & --- & 2.8$\to$20.0 & 2009--19~\cite{hmv_poland2022} \\
Hungary & 384 & 40 (10.4\%) & 3.9 & 2018~\cite{hmv_hungary2018} \\
South Korea & --- & 62.8\% trach. & 9.3 & 2016~\cite{hmv_korea2019} \\
Germany & $\sim$17{,}000/yr$^{\dagger}$ & $\sim$6\% & --- & 2018~\cite{hmv_germany2021} \\
USA & no registry$^{\ddagger}$ & --- & --- & ---~\cite{us_trach_mehta2015} \\
\bottomrule
\end{tabular}

\smallskip
{\footnotesize \emph{Note.} Metric types differ: point-prevalence headcount (Japan),
prevalence per 100{,}000 (Europe, Canada, Poland, Hungary, South Korea),
$^{\dagger}$inpatient \emph{episodes/year} not living prevalence (Germany), and
$^{\ddagger}$no national home-ventilation registry in the USA (hospital data show home
discharge of tracheostomy-ventilation patients declining). Not strictly comparable.}
\end{table}

What makes bFaaaP relevant to these groups is its \emph{controller} rather than the pedal.
The controller is a commodity-smartphone, quantitative, user-tunable head-angle channel
that (i)~is \emph{foot-free}, so it does not depend on the lower limbs that wheelchair users
often cannot use; (ii)~attaches \emph{nothing to the face or head}---the phone sits on a
stand---which matters for people with a tracheostomy, facial sensitivity, or who must keep
the airway and face clear; and (iii)~is \emph{tunable to a restricted range of motion}: as
the Class~III participant showed, a small offset with a large multiplier lets a few degrees
of residual head movement span the full output (Section~\ref{sec:impact}). These are
precisely the constraints the two populations above present. Because the head-angle signal
is a continuous, pre-set, \emph{proportional} value---not a single on/off switch---it is a
general \emph{accessibility-control} primitive: the same channel that here meters a sustain
pedal could, in principle, meter other graded controls (an environmental-control setting, a
communication-aid scan rate, a powered-device level). The device-controller method is
patented independently of the pedal and explicitly spans ``any device''~\cite{jp7004771},
and the \emph{Switch} line already shows how cheaply the controller retargets to a new
actuator. We present this as a discussion-level argument and a direction for future work:
bFaaaP \emph{validates} the channel for expressive piano pedalling, and its tunable,
face-free, foot-free design suggests the same controller could contribute to accessibility
control for the much larger populations of Tables~\ref{tab:wheelchair}
and~\ref{tab:hmv}---which we have not yet validated and leave to future work.

\subsection{Comparison with existing systems}
Table~\ref{tab:comparison} compares bFaaaP with representative prior systems on a
literature basis. Relative to Steingraeber's instrument-integrated electromagnetic
control~\cite{steingraeber,mouthpedal2008} and CanAssist's head-activated
pedal~\cite{canassist}, bFaaaP is distinguished by a commodity-smartphone sensor,
a quantitative \emph{user-tunable} (rather than binary) law, \emph{non-destructive}
placement on an unmodified piano, explicit \emph{clinical} validation, and open
source release. Its main present limitations are a dependence on a clear view of
the face and per-instrument calibration (discussed below).

\begin{table}[tbp]
\caption{Comparison with representative assistive piano-pedal approaches
(literature-based; n.r.\ = not reported).}
\label{tab:comparison}
\centering
\small
\setlength{\tabcolsep}{4pt}
\newcolumntype{L}[1]{>{\raggedright\arraybackslash}p{#1}}
\begin{tabular}{@{}L{2.6cm}L{2.4cm}L{2.0cm}L{2.6cm}L{2.0cm}L{1.6cm}@{}}
\toprule
System & Input & Non-destructive & User-tunable & Commodity sensor & Open source \\
\midrule
\textbf{bFaaaP} (Pro/Switch) & head angle, smartphone AR & yes (airback) & yes (offset + mult. $\rightarrow$ speed) & yes & yes \\
Steingraeber \cite{steingraeber,mouthpedal2008} & helmet tilt / neck reed / bite force & no (integrated) & partial (bite) & no & no \\
CanAssist~\cite{canassist} & head tilt, headband & yes (floor unit) & no (binary) & no & no \\
Electromech.\ activators \cite{hinsley1988,uspedal2008,uspedal2017} & varied & varies & typically binary & no & no \\
\bottomrule
\end{tabular}
\end{table}

\subsection{Non-destructive anchoring and portability (airback)}
A practical contribution is the pneumatic \emph{airback}
(Section~\ref{sec:airback}). Pressing a pedal creates a reaction force that would
push a light device away; rather than bolt the unit to a valuable, shared
instrument, bFaaaP inflates a bag under a neighbouring pedal to absorb that force.
The result is a device that is lightweight, mechanically simple, strongly anchored,
and quick to set up and remove---so it suits not only home practice but
\emph{concert} use on instruments that must not be modified. This portability is a
meaningful accessibility property: it lets a player bring their pedalling to many
venues rather than to one adapted piano.

\subsection{Generality and the practical value of the controller}
\label{sec:generality}
The smartphone controller is, we argue, the component of highest practical value.
It is a quantitative, user-tunable head-angle channel built on commodity hardware,
and nothing about it is specific to a pedal: the same controller already drives two
different actuators (a motor and an electronic switch), and the underlying
device-controller method is patented independently of the pedal
application~\cite{jp7004771}, covering electronic instruments and, more broadly,
``any device.'' This generality matters in two ways. First, the \emph{Switch} line
shows how cheaply the controller scales---an electronic sustain needs no
drivetrain---so head-angle pedalling can reach the very large installed base of
digital pianos and keyboards at low cost. Second, the controller is a reusable
\emph{inclusive input}: the tunable head channel could operate
other expressive controls, other instruments, or assistive devices beyond music.
We therefore see bFaaaP not only as a pedal aid but as a platform for inclusive,
hands-free control of a wide range of products and services.

\subsection{Future perspectives}
\label{sec:future}
The co-author who designed and built the pedal device (H.N.) frames the open-source
release as a deliberate \emph{stepping stone}: the project is not finished, and
opening it invites a wider community to improve it, lower its cost, and find
approaches the original team would not. A useful way to organize that work is to
decompose the foot-free assistive pedal into three largely independent problems:
(i)~how to \emph{sense} a non-foot body movement that carries the player's intention;
(ii)~how to \emph{transmit} that signal to the actuator quickly and reliably; and
(iii)~how to \emph{actuate} the pedal quickly and quietly. Each axis admits
substitution without disturbing the others---which is exactly what the
controller/actuator split already demonstrates (Section~\ref{sec:generality}).
Concrete directions follow. On sensing, a dedicated low-cost sensor in place of a
smartphone could reduce total cost and broaden reach, and supporting Android
alongside iOS would widen access. On transmission, the rate-decoupled BLE channel
(Section~\ref{sec:decoupling}) is already device-agnostic. On actuation, the drive
need not remain a stepper and lead screw: the same control law could meter a
closed-loop servo-stepper today and, in principle, softer or more compact actuators
such as artificial-muscle or piezoelectric drives, extending the method beyond the
piano. We present these as directions rather than results; subsequent iterations
will be refined openly with the team and the broader community of contributors.

\subsection{Limitations and future work}
bFaaaP depends on a clear view of the player's face and on per-instrument
calibration, and the present quantitative evidence is the APEE re-analysis plus
field deployments rather than a controlled, pre-registered user study with latency
and usability instruments; such a study (including more participants with
disabilities) is planned. The reference Pro design published here uses an IQ/Fortiq M42BLS
smart-servo motor; because that motor is now end of life, a closed-loop NEMA17-class
stepping motor with load feedback is under evaluation as its successor with the
device co-author, but that firmware is still early and is not yet the reference, as
the open hardware documentation tracks. The Switch line invites its own formal evaluation;
and while a single-sheet KiCad schematic of the IQ reference generation is already
available in the repository, a stepper-version schematic and a build-clean
stepper-firmware reference are still being finalized with the device co-author.
Planned next-generation simplifications include a single 24\,V supply with
on-board 5\,V regulation, and an airback that stays closed with only a release
mechanism (and possibly pump-current rather than a pressure sensor) to make setup
faster and more robust.

% --- 5. Implementation Procedure (drafted to sub-subsection level) ---
% =====================================================================
% Draft section for 作業7 (paper full revision) — §5
% "Implementation Procedure" — instruction-style, to sub-subsection level.
% Compiles in BOTH arXiv (article) and TACCESS (acmart). Cites: repo,
% rp2040, arkit. Cross-refs: sec:realization, sec:clinical, sec:airback,
% tab:participants (APEE participant table in §3.3).
% Honest reproducibility note: a consolidated wiring schematic and a
% build-clean main firmware are being finalized with the device co-author;
% the repository's hardware notes track the remaining gaps.
% =====================================================================

\section{Implementation Procedure}
\label{sec:impl}

This section is written as a reproducibility guide: all software is open source
and the hardware design files (CAD, print files, firmware sketches, bill of
materials) are published in the project repository~\cite{repo}, so that a reader
can rebuild the shared iOS controller, the \emph{Pro} acoustic device (the
primary, hardware-intensive build), and the \emph{Switch} electronic device, and
re-run the APEE-style evaluation of Section~\ref{sec:clinical}. Steps reference
the realized architecture of Section~\ref{sec:realization}; paths below are
folders in the repository.

\subsection{Open-source repository and prerequisites}
\label{sec:impl-repo}

\subsubsection{Repository layout}
\label{sec:impl-layout}
The repository~\cite{repo} separates the shared controller from the two device
lines: \texttt{ios-app/} (app source plus \texttt{CODE-STRUCTURE},
\texttt{DESIGN-HIGHLIGHTS}, \texttt{BLE-CONNECTION-FLOW}, and \texttt{BUILD}
notes); \texttt{device-pro-acoustic/} (\texttt{firmware/}, \texttt{hardware/}
CAD and 3D-print files, \texttt{motor/}, \texttt{assembly/}, and a hardware
availability note); \texttt{device-switch-electronic/}; shared \texttt{docs/}
(\texttt{architecture/}, \texttt{operation/}, \texttt{toolchain/},
\texttt{user-manual/}); and \texttt{bfaaap\_patent\_info/}. Code carries no
personal signing identifiers, so a new developer supplies their own team and
bundle identifier.

\subsubsection{Toolchain prerequisites}
\label{sec:impl-toolchain}
The controller is built with Xcode and requires an Apple developer team and a
TrueDepth (Face~ID-class) iPhone or iPad, because ARKit face tracking is
device-only~\cite{arkit}. The device firmware is built from VS~Code with
PlatformIO (or the Arduino extension), the Adafruit nRF52 board support package
for the BLE board and the \texttt{arduino-pico} core for the RP2040 main
board~\cite{rp2040}. The 3D-printed parts are produced with any slicer; the
\texttt{docs/toolchain/} guide gives flashing and slicing details.

\subsection{Building the iOS controller (shared by Pro and Switch)}
\label{sec:impl-ios}

\subsubsection{Configure signing and build}
\label{sec:impl-ios-build}
Clone the repository, open \texttt{ios-app/src}, set \texttt{DEVELOPMENT\_TEAM}
and the bundle identifier to your own (placeholders are provided), keep automatic
signing, and build to a connected TrueDepth device. The \texttt{ios-app/BUILD}
note lists the exact steps; no proprietary credentials are needed.

\subsubsection{Run and verify AR$\rightarrow$BLE operation}
\label{sec:impl-ios-verify}
On first launch, grant camera permission; the play screen shows the head-tracking
state and turns red once the head passes the threshold. With a powered device
nearby, the app discovers and name-filters \texttt{bFaaaPSwitch\_n}, connects to
the selected channel, and begins streaming once the BLE write characteristic is
ready.

\subsubsection{Individualized pre-setting (offset and multiplier)}
\label{sec:impl-preset}
Before playing, the player performs a one-time \emph{individualized pre-setting}
step in the app: they choose the threshold start-point (offset) and the multiplier
to match their own comfortable range and pace of head motion, and the app stores
this preset. This step is what turns the control law of
Section~\ref{sec:control-law} into \emph{individualized control}---the same device
serves an able-footed adult, a child, and a player with restricted head motion,
each with their own offset/multiplier (e.g.\ the tracheostomy participant of
Section~\ref{sec:impact} chose a small offset and a large multiplier). Treat this
preset as a required deployment step for each new player.

\subsection{Building the Pro device (acoustic; primary)}
\label{sec:impl-pro}

\subsubsection{Mechanical assembly and 3D-printed parts}
\label{sec:impl-pro-mech}
The chassis is a \emph{vertical} aluminium-extrusion frame---a 2040 post on a 2080
base, with 4040 stock---carrying 3D-printed housings (the reference parts were
printed on a 305\,mm-class printer in PLA+); the largest part needs a bed of about
$\ge$300\,mm and $\sim$200\,mm of height. The drivetrain is
motor~$\rightarrow$ timing belt $\rightarrow$ lead screw $\rightarrow$ push-rod
(Section~\ref{sec:pro-chain}): a GT-2 ($262$\,mm) timing belt and $T60/T60$ pulleys
(a $1{:}1$ ratio) turn a vertical $T10$ lead screw ($16$\,mm/rev) whose carriage
drives the push-rod straight down onto the sustain pedal. The CAD, print files, and
the full bill of materials are in \texttt{device-pro-acoustic/hardware/}.

\subsubsection{Electronics, motor drive, and power}
\label{sec:impl-pro-elec}
Two boards are used: an nRF52840-class BLE board (an AE-NRF52840) and an RP2040
main board joined by UART. The \emph{reference} design published here uses an
\emph{IQ/Fortiq M42BLS} smart-servo motor driven over a serial interface; because
that motor is now end of life, the \emph{next} version will use a closed-loop
\emph{NEMA17}-class stepping motor---an integrated closed-loop servo-stepper that
reports load, so the force feedback the IQ servo provided is preserved (the
specific model is under evaluation with the device co-author)---but that firmware
is still in development and is not the reference here (the hardware-availability
note documents the planned substitution). Pedal reaction is sensed from
\emph{motor power} rather than current, because power is robust to supply-voltage
variation (on the IQ version the power is read by command from the motor); a
commercial DC supply sized for the motor is recommended. A \emph{single-sheet KiCad
schematic} of the reference (IQ) generation --- giving the full board-to-board
wiring and the Pico pin map (BLE UART on GP0/GP1, the HX711 airback-pressure sensor on GP2/GP3, the
IQ-motor serial on GP4/GP5, the air-pump MOSFET on GP12, and the slider
potentiometer on ADC0/GP26) and two power rails (+5\,V logic, +24\,V motor) ---
is included in the repository~\cite{repo}; a stepper-version schematic and a
build-clean stepper firmware are still being finalized with the device co-author.

\subsubsection{Airback (reaction anchoring) and hand controller}
\label{sec:impl-pro-air}
The airback is a commercial air-wedge cushion (e.g.\ a WINBAG). Wire its electric
air pump through a power MOSFET (GP12) to the controller, plumb it to the cushion
placed under a neighbouring pedal, and connect a small hand controller (an
upper-limit \emph{slide potentiometer} read on ADC0/GP26, plus a pump on/off
switch); the built device drives the pump electrically, with the cushion's manual
hand-bulb as a no-electronics alternative. At power-up the firmware inflates the
airback until the pressure threshold anchors the device, then self-calibrates the
lower travel end from the motor-power limit (a DIP switch sets the pressing force,
$\sim$20--35\,W, which is piano-dependent; Section~\ref{sec:airback}).

\subsubsection{Firmware flashing and device naming}
\label{sec:impl-pro-fw}
Flash the BLE board through its UF2 bootloader (Adafruit nRF52 workflow) and the
RP2040 through BOOTSEL with the \texttt{arduino-pico} core, following
\texttt{docs/toolchain/}. The advertised name (\texttt{bFaaaPSwitch\_1}..\texttt{\_4})
is stored in the BLE board's flash so several units can coexist on distinct
channels.

\subsubsection{On-piano setup and self-calibration}
\label{sec:impl-pro-setup}
Place the device on a board spanning the pedals, align the push-rod over the
sustain (damper) pedal, inflate the airback until the unit is firmly anchored,
let the actuator auto-find the lower end against the pedal, set the upper end with
the slider, and pick the on-type/off-type to match the instrument
(\texttt{docs/operation/}). The setup is non-destructive and portable, which is
what makes concert use --- not only home use --- practical.

\subsection{Building the Switch device (electronic)}
\label{sec:impl-switch}

\subsubsection{Electronics and sustain-jack interface}
\label{sec:impl-switch-elec}
The Switch needs only an nRF52840-class board (an Adafruit ItsyBitsy nRF52840) and
a logic-level N-channel MOSFET as a low-side switch (e.g.\ a ROHM RU1J002YN) from a
GPIO pin to the instrument's sustain jack (a 6.3\,mm plug); there is no motor,
drivetrain, or airback. It runs from $2\times$AA cells and starts on reset, and an
RGB LED provides the idle/armed/engaged color feedback.

\subsubsection{Setup}
\label{sec:impl-switch-setup}
Connect the lead to the digital piano's sustain jack, pair the channel in the
app, and set the on-type/off-type to match the jack polarity. Operation then
follows the same head-angle control law as the Pro device.

\subsection{Reproducing the APEE evaluation}
\label{sec:impl-apee}

\subsubsection{Test protocol}
\label{sec:impl-apee-protocol}
Recruit participants by class (I: able to use their own feet, as control; II:
children; III: users with leg/neck disabilities). Each participant practices the
reference score (the motif do--do--do, re--re--re, mi--mi--mi repeated four
times), selects an offset and multiplier to taste, and records three conditions
of the \emph{same} motif: \emph{pattern~0}, played with no sustain pedal as the
baseline; and \emph{pattern~1} and \emph{pattern~2}, played with bFaaaP under two
prescribed sustain-pedalling schemes. The two schemes are marked on the score as
pedal-down spans drawn under the staff (one row of spans per pattern): pattern~1
changes the pedal at each three-note group, whereas pattern~2 holds it across the
groups for a longer, more connected sustain. The two therefore probe
\emph{controllability} rather than a single on/off effect; in the original study
pattern~2 was the more sustained and scored higher than pattern~1
(Section~\ref{sec:clinical}). A short questionnaire follows (the participant
table is Table~\ref{tab:participants}).

\subsubsection{Analysis and statistics}
\label{sec:impl-apee-analysis}
Import each recording into an audio analyzer (Sonic Visualiser), export the
note regions as images, and measure a tone-vibration area (TVA) with ImageJ's
particle analysis. Normalize the no-pedal condition to $1.00$ and express
patterns~1 and~2 as ratios, then compare conditions with paired $t$-tests and
compare across classes, piano experience, and instruments with one-way ANOVA.
This reproduces the result tables of Section~\ref{sec:clinical}, in which bFaaaP
significantly increased sustain ($p<0.01$) and was statistically
indistinguishable from a participant's own foot ($p>0.05$).

% =====================================================================
\section{Conclusion}
\label{sec:conclusion}
bFaaaP turns a small, natural head movement into expressive, foot-quality
piano-pedal control. Its inventive core is a quantitative, user-tunable control
law---an angular dead-zone and a user multiplier that the player presets, which
together fix a secondary, temporal \emph{response speed} (how fast the pedal
actuator follows the head past the dead-zone)---validated in a human-subject study
to be statistically equivalent to a
player's own foot and effective across adults, children, and people with
disabilities, including a participant with a tracheostomy. One commodity-smartphone
controller drives two co-equal realizations: a non-destructive robotic actuator
for acoustic pianos and an electronic switch for digital instruments. By
open-sourcing the full system---and offering the patented method free of charge for
inclusive uses---we aim to make pedal-based piano performance inclusive: available
to anyone, regardless of whether they can use a foot.

% =====================================================================
\section*{Acknowledgements}
We thank Yasuhiro Ono for his advice on AI programming, and Taguchi for
contributing the KiCad schematic and electronics of the Pro device. We are grateful to the
Institute of Science Tokyo and to the owner couple of MUSICASA
(\url{https://www.musicasa.co.jp}) for kindly providing venues for our
demonstrations and concerts. We sincerely thank the participants of the Auxiliary
Pedal Effect Evaluation (APEE)---the adults, children, and players with
disabilities, and their families---for generously taking part and sharing their
feedback. We also thank the bFaaaP and Platanus members and
collaborators, and the performers and audiences of the demonstration events.
Caricature illustrations of the team are by Saki Shiokawa.

\section*{Funding}
This bFaaaP project was supported by Shishido \& Associates.

\section*{Ethics}
The Auxiliary Pedal Effect Evaluation (APEE) involved human participants: adults,
children aged 7--11, and people with disabilities. Participation was voluntary, and
informed consent was obtained in writing for every participant. Adult participants gave
their own written informed consent. For the children, a parent or guardian confirmed
written informed consent (arranged through a piano teacher who assisted the study), and
the child signed the consent form themselves. Participants with disabilities took part
with the written consent of a parent or guardian, who also accompanied them and provided
transport to and from the sessions. Any participant could stop or withdraw at any time,
and the tasks (playing short passages on a piano) posed no more than minimal risk. The
authors are independent practitioners and are not affiliated with an institution that
operates an institutional review board (IRB); no formal IRB review or approval number is
therefore available. The study was nonetheless conducted following the ethical principles
of the ACM Publications Policy on Research Involving Human Participants and
Subjects---minimizing potential harm and protecting participants' privacy and right to
self-determination. All reported data are anonymized (Appendix~A), and no personally
identifying information is included.

\section*{Conflict of interest/Competing interests}
T.S.\ is a chief technology officer at SeemeData Labs, Inc.\ and chief Japanese
patent attorney at Shishido \& Associates. T.S., M.O., and H.N.\ are among the
inventors on the patents underlying bFaaaP (JP~6726319 and JP~7004771;
PCT~WO~2019/176164), which are held by T.~Shishido and Ootaki \& Architects, and
are involved in its commercialization (the \emph{bFaaaPSwitch} iOS application and
the build-to-order pedal device); \emph{bFaaaP} is a registered trademark. The
authors otherwise declare no competing interests.

\section*{Patent availability}
The control method is protected by JP~6726319~B2 and JP~7004771~B2 (priority basis:
PCT~WO~2019/176164~\cite{pct2019}). As a matter of policy the authors intend to license the
patents \emph{free of charge} for genuinely public/open uses and, in particular,
for products or services that enable inclusive social participation by people with
disabilities---even commercial ones. Please contact the corresponding author.

\section*{Data and code availability}
The iOS application source, device firmware, mechanical (CAD/3D-print) designs,
documentation, and the patent file-wrapper materials underlying the APEE data are
openly available in the project repository~\cite{repo}. The intended release
licensing (final decision pending) is a three-layer scheme: \emph{software}
(iOS app and firmware) under the \textbf{Apache License 2.0} (which carries an
explicit patent grant), \emph{hardware} designs (CAD/3D-print/schematic) under
\textbf{CERN-OHL-W-2.0}, and \emph{documentation} under \textbf{CC-BY-4.0};
practising the patented invention in a device is governed by the separate patent
policy above.

% =====================================================================
% =====================================================================
% Appendix: full anonymised APEE per-recording data (No.1-15, 46 recordings).
% Transcribed from the original PCT/patent study figures (anonymised: subjects
% appear only as No.1-15). Pattern 0 (no pedal) is normalised to 1.00 for every
% recording, so only the Pattern 1 / Pattern 2 relative TVA are tabulated.
% Requires \usepackage{longtable} (added in main.tex).
% =====================================================================
\appendix
\section{Full APEE per-recording data (anonymised)}
\label{app:apee-data}

Table~\ref{tab:apee-full} lists every Auxiliary Pedal Effect Evaluation (APEE)
recording (Section~\ref{sec:clinical}), transcribed from the original study
figures. Subjects are anonymised as \textbf{No.~1--15}. ``P1''/``P2'' are the
relative tone-vibration area (TVA) of pedalling pattern~1 and pattern~2; the
no-pedal baseline (pattern~0) is $1.00$ for every recording and is omitted.
``Reg.\ pedal'' rows (Class~I only) are the player's own foot, used as a control.
The original anonymised figures are archived in the
repository.\footnote{\texttt{docs/history/pct-original-figures/} (files
\texttt{apee-04-...} and \texttt{apee-05-...}).}

\footnotesize
\setlength{\tabcolsep}{4pt}
\begin{longtable}{@{}r l l l l r r r r r@{}}
\caption{APEE per-recording data, all 46 recordings, anonymised (No.~1--15).
P1/P2 = relative TVA (pattern~0 $\equiv 1.00$). Off.\ = offset (deg); Mult.\ =
multiplier; n/a = own foot / not applicable; N.D.\ = not recorded.}
\label{tab:apee-full}\\
\toprule
No. & Age & Exp. & Piano & Device & Off. & Mult. & P1 & P2 & \# \\
\midrule
\endfirsthead
\multicolumn{10}{@{}l}{\emph{(Table~\ref{tab:apee-full} continued)}}\\
\toprule
No. & Age & Exp. & Piano & Device & Off. & Mult. & P1 & P2 & \# \\
\midrule
\endhead
\midrule
\multicolumn{10}{@{}r}{\emph{continued on next page}}\\
\endfoot
\bottomrule
\endlastfoot
\multicolumn{10}{@{}l}{\textbf{Class~I --- adults (controls; can use own feet)}}\\
1 & 50 & 1--2y (child) & K132 & Reg.\ pedal & n/a & n/a & 1.26 & 1.81 & 1 \\
  &    &               & K132 & bFaaaP2 (M5) & 5 & 15 & 1.05 & 1.26 & 2 \\
  &    &               & K132 & bFaaaP3 & N.D. & N.D. & 1.15 & 1.27 & 3 \\
2 & 66 & 0y & K132 & bFaaaP4 & 19 & 29 & 1.86 & 3.30 & 4 \\
3 & 53 & 0y & K132 & bFaaaP4 & 10 & 30 & 1.41 & 1.78 & 5 \\
  &    &    & UX   & Reg.\ pedal & n/a & n/a & 1.29 & 1.31 & 6 \\
  &    &    & UX   & bFaaaP4 & 15 & 25 & 1.30 & 1.58 & 7 \\
  &    &    & UX   & bFaaaP4 & 10 & 40 & 1.25 & 1.45 & 8 \\
  &    &    & K132 & Reg.\ pedal & n/a & n/a & 1.29 & 1.51 & 9 \\
  &    &    & GC1  & bFaaaP4W & 5 & 20 & 1.67 & 1.97 & 10 \\
  &    &    & GC1  & bFaaaP4W & 5 & 30 & 1.70 & 1.78 & 11 \\
  &    &    & GC1  & Reg.\ pedal & n/a & n/a & 1.41 & 1.41 & 12 \\
4 & 48 & many y & K132 & bFaaaP4 & 5 & 20 & 1.77 & 1.97 & 13 \\
  &    &        & K132 & bFaaaP4 & 10 & 20 & 1.63 & 1.70 & 14 \\
  &    &        & K132 & bFaaaP4 & 5 & 40 & 1.80 & 1.94 & 15 \\
  &    &        & K132 & Reg.\ pedal & n/a & n/a & 1.90 & 2.46 & 16 \\
5 & 35 & many y & K132 & bFaaaP4 & 10 & 30 & 1.64 & 1.83 & 17 \\
  &    &        & K132 & bFaaaP4 & 15 & 30 & 1.73 & 1.83 & 18 \\
  &    &        & K132 & Reg.\ pedal & n/a & n/a & 1.92 & 2.06 & 19 \\
  &    &        & GC1  & bFaaaP4W & 5 & 20 & 1.94 & 1.97 & 20 \\
  &    &        & GC1  & bFaaaP4W & 5 & 30 & 1.82 & 1.93 & 21 \\
  &    &        & GC1  & Reg.\ pedal & n/a & n/a & 1.88 & 2.20 & 22 \\
6 & teens & 3y & UX & Reg.\ pedal & n/a & n/a & 1.18 & 1.45 & 23 \\
  &       &    & UX & bFaaaP4 & 5 & 10 & 1.43 & 1.48 & 24 \\
  &       &    & UX & bFaaaP4 & 10 & 30 & 1.19 & 1.28 & 25 \\
7 & forties & 6y & UX & Reg.\ pedal & n/a & n/a & 1.08 & 1.06 & 26 \\
  &         &    & UX & bFaaaP4 & 10 & 30 & 1.27 & 1.27 & 27 \\
  &         &    & UX & bFaaaP4 & 5 & 10 & 1.12 & 1.17 & 28 \\
\midrule
\multicolumn{10}{@{}l}{\textbf{Class~II --- children (feet do not reach the pedals)}}\\
8  & 11 & 2y7m & GC1 & bFaaaP4W & 5 & 20 & 1.83 & 2.19 & 29 \\
   &    &      & GC1 & bFaaaP4W & 5 & 30 & 2.05 & 2.23 & 30 \\
9  & 8  & 1y6m & GC1 & bFaaaP4W & 5 & 20 & 1.87 & 1.78 & 31 \\
   &    &      & GC1 & bFaaaP4W & 5 & 30 & 1.19 & 1.01 & 32 \\
10 & 10 & 5y   & GC1 & bFaaaP4W & 5 & 20 & 2.17 & 2.42 & 33 \\
   &    &      & GC1 & bFaaaP4W & 5 & 30 & 1.66 & 1.75 & 34 \\
11 & 8  & 2y   & GC1 & bFaaaP4W & 5 & 20 & 2.18 & 2.57 & 35 \\
   &    &      & GC1 & bFaaaP4W & 5 & 30 & 1.56 & 1.56 & 36 \\
12 & 7  & 2y   & GC1 & bFaaaP4W & 5 & 20 & 2.06 & 1.89 & 37 \\
   &    &      & GC1 & bFaaaP4W & 5 & 30 & 1.23 & 1.68 & 38 \\
\midrule
\multicolumn{10}{@{}l}{\textbf{Class~III --- people with disabilities}}\\
13 & 16 & 0y & K132 & bFaaaP2 (M5) & 10 & 15 & 1.09 & 0.90 & 39 \\
   &    &    & K132 & bFaaaP3 & N.D. & N.D. & 1.32 & 1.30 & 40 \\
14 & 19 & several y & K132 & bFaaaP4 & 5 & 40 & 1.18 & 1.36 & 41 \\
   &    &          & K132 & bFaaaP4 & 3 & 50 (max) & 0.98 & 1.67 & 42 \\
15 & 50 & 5y & K132 & bFaaaP2 (M5) & 5 & 8 & 1.72 & 2.12 & 43 \\
   &    &    & K132 & bFaaaP4 & 10 & 30 & 1.84 & 2.22 & 44 \\
   &    &    & K132 & bFaaaP4 & 10 & 40 & 1.70 & 1.79 & 45 \\
   &    &    & K132 & bFaaaP4 & 5 & 40 & 1.56 & 2.17 & 46 \\
\end{longtable}
\normalsize

\noindent\emph{Notes.} No.~13 and No.~15 have a leg disability; No.~14 has a
partial leg disability \emph{and} a tracheotomy (and chose a small offset with a
large multiplier --- recordings 41--42). Offsets for the early ``bFaaaP2 (M5)''
(M5Stack) generation are reference values only (the sensor sat in a cap pocket);
``bFaaaP3'' kept no parameter memory, so its offset/multiplier are not recorded
(N.D.). Piano models: K132 (Steinway), GC1 (Yamaha), and UX (Yamaha).

% =====================================================================
\bibliographystyle{unsrt}
\bibliography{refs}

\begin{thebibliography}{10}

\bibitem{frid2019}
Emma Frid.
\newblock Accessible digital musical instruments---a review of musical
  interfaces in inclusive music practice.
\newblock {\em Multimodal Technologies and Interaction}, 3(3):57, 2019.

\bibitem{ramirez2023}
Rafael Ramirez-Melendez.
\newblock Accessible digital music instruments for motor disability.
\newblock In {\em Neurocognitive Music Therapy}. Springer, 2023.

\bibitem{duarte2023}
Erivan Gon{\c{c}}alves~Duarte, Isabelle Cossette, and Marcelo~M. Wanderley.
\newblock Analysis of accessible digital musical instruments through the lens
  of disability models: a case study with instruments targeting d/{D}eaf
  people.
\newblock {\em Frontiers in Computer Science}, 5, 2023.

\bibitem{eyeharp}
Zacharias Vamvakousis and Rafael Ramirez.
\newblock The {EyeHarp}: A gaze-controlled digital musical instrument.
\newblock {\em Frontiers in Psychology}, 7:906, 2016.

\bibitem{sumak2019}
Bo{\v{s}}tjan {\v{S}}umak, Matic {\v{S}}pindler, Mojca Debeljak, Marjan
  Heri{\v{c}}ko, and Maja Pu{\v{s}}nik.
\newblock An empirical evaluation of a hands-free computer interaction for
  users with motor disabilities.
\newblock {\em Journal of Biomedical Informatics}, 96:103249, 2019.

\bibitem{varona2008}
Javier Varona, Cristina Manresa-Yee, and Francisco~J. Perales.
\newblock Hands-free vision-based interface for computer accessibility.
\newblock {\em Journal of Network and Computer Applications}, 31(4):357--374,
  2008.

\bibitem{manresa2014}
Cristina Manresa-Yee, Javier Varona, Francisco~J. Perales, and Iosune Salinas.
\newblock Design recommendations for camera-based head-controlled interfaces
  that replace the mouse for motion-impaired users.
\newblock {\em Universal Access in the Information Society}, 13(4):471--482,
  2014.

\bibitem{arkit}
{Apple Inc.}
\newblock {ARKit}: Tracking and visualizing faces (developer documentation).
\newblock \url{https://developer.apple.com/documentation/arkit}.
\newblock Accessed 2026.

\bibitem{weinberg2020}
Gil Weinberg, Mason Bretan, Guy Hoffman, and Scott Driscoll.
\newblock {\em Robotic Musicianship: Embodied Artificial Creativity and
  Mechatronic Musical Expression}.
\newblock Springer, 2020.

\bibitem{scimeca2020}
Luca Scimeca, Cheryn Ng, and Fumiya Iida.
\newblock Gaussian process inference modelling of dynamic robot control for
  expressive piano playing.
\newblock {\em PLOS ONE}, 15(8):e0237826, 2020.

\bibitem{zappi2012}
Victor Zappi, Antonio Pistillo, Sylvain Calinon, Andrea Brogni, and Darwin~G.
  Caldwell.
\newblock Music expression with a robot manipulator used as a bidirectional
  tangible interface.
\newblock {\em EURASIP Journal on Audio, Speech, and Music Processing}, 2012:2,
  2012.

\bibitem{canassist}
{CanAssist, University of Victoria}.
\newblock Head-activated piano pedal.
\newblock
  \url{https://web.archive.org/web/20180402202115/https://www.canassist.ca/EN/main/programs/technologies-and-devices/test-1/piano-arts.html},
  2018.
\newblock Accessed via Internet Archive (capture 2 Apr 2018).

\bibitem{steingraeber}
{Steingraeber \& S{\"o}hne}.
\newblock Pedal devices for pianists in wheelchairs.
\newblock
  \url{https://www.steingraeber.de/en/innovationen/pedal-devices-for-pianists-in-wheelchairs/}.
\newblock Accessed 2026.

\bibitem{mouthpedal2008}
{ScienceDaily}.
\newblock Paraplegic pianists can operate a piano pedal with the mouth.
\newblock \url{https://www.sciencedaily.com/releases/2008/10/081024103211.htm},
  2008.
\newblock Article dated 24 Oct 2008. Accessed 2026.

\bibitem{hinsley1988}
J.~D. Hinsley and Bobby Norwood.
\newblock Piano pedal activator for paraplegics, 1988.
\newblock U.S. Patent 4{,}736{,}664.

\bibitem{uspedal2008}
Pedaling aid for handicapped musician, 2008.
\newblock U.S. Patent 7{,}432{,}429.

\bibitem{uspedal2017}
Piano pedal operating device for people with disabilities, 2017.
\newblock U.S. Patent 9{,}792{,}885.

\bibitem{piano_manipulator2025}
Pu-Sheng Tsai, Ter-Feng Wu, and Chen-Ting Liao.
\newblock Development of a robotic manipulator for piano performance via
  numbered musical notation recognition.
\newblock {\em Machines}, 13(12):1121, 2025.

\bibitem{ctf_actuator2020}
Manmatha Mahato, Rassoul Tabassian, Van~Hiep Nguyen, Saewoong Oh, Sanghee Nam,
  Won-Jun Hwang, and Il-Kwon Oh.
\newblock {CTF}-based soft touch actuator for playing electronic piano.
\newblock Nature Communications 11, article 5358, 2020.

\bibitem{hmd_jp2017}
{Sony Interactive Entertainment Inc.}
\newblock Operation input device and operation input method, 2017.
\newblock Japan Patent Application Publication JP2017-21461A (cited as ref. 1
  in the JP6726319/JP7004771 examinations; a head-movement input device with an
  on-screen indicator of whether a posture-change threshold has been crossed).

\bibitem{headset_jp2012}
Control system, earphone, and control method, 2012.
\newblock Japan Patent (Tokuhyo) Publication JP2012-514392A (cited as ref. 1 in
  the JP7004771 appeal).

\bibitem{colopl_jp2017}
{COLOPL, Inc.}
\newblock Method implemented by a head-mounted display system, program, and
  head-mounted display system, 2017.
\newblock Japan Patent Application Publication JP2017-37567A (ISR D2).

\bibitem{titech_jp2014}
{Tokyo Institute of Technology}.
\newblock Operation system for operation-object device and operation input
  device, 2014.
\newblock Japan Patent Application Publication JP2014-95953A; family WO
  2014/073121 A1 (ISR D3).

\bibitem{alps_us2018}
{Alps Electric Co., Ltd.}
\newblock Posture detection device, spectacle-type electronic device, posture
  detection method, and program, 2018.
\newblock U.S. Patent Application Publication US 2018/0064371 A1; family WO
  2016/194581 A1, EP 3305193 A1 (ISR D4).

\bibitem{yamaha_jp2006}
{Yamaha Corporation}.
\newblock Pedal performance assisting device, 2006.
\newblock Japan Patent Application Publication JP2006-154504A; family US
  2006/0112809 A1 (ISR D5).

\bibitem{jp6726319}
Tomoyuki Shishido, Hiroyuki Narusawa, Masahiro Ootaki, Kyoko Yamaguchi, and
  Daisuke Tokushige.
\newblock Auxiliary pedal system, 2020.
\newblock Japan Patent JP6726319B2, registered 30 June 2020 (application
  2018-567754; priority 12 Nov 2018).

\bibitem{jp7004771}
Tomoyuki Shishido, Hiroyuki Narusawa, Masahiro Ootaki, Kyoko Yamaguchi, and
  Daisuke Tokushige.
\newblock Device controller, 2022.
\newblock Japan Patent JP7004771B2, registered 6 Jan 2022 (divisional
  application 2020-110051).

\bibitem{hmv_japan2020}
{Japan Ministry of Health, Labour and Welfare (Intractable-Disease Policy
  Research)}.
\newblock Nationwide survey of home mechanical-ventilation users in japan, by
  prefecture.
\newblock \url{https://mhlw-grants.niph.go.jp/}, 2021.
\newblock As of 31 Mar 2020: ca.\ 7{,}700 home tracheostomy positive-pressure
  ventilation (TPPV) users and ca.\ 13{,}500 noninvasive (NPPV) users (ca.\
  21{,}000 total). MHLW Intractable-Disease Policy Research report. Accessed
  2026.

\bibitem{eurovent2005}
S.~J. Lloyd-Owen, G.~C. Donaldson, N.~Ambrosino, J.~Escarrabill, R.~Farre,
  B.~Fauroux, D.~Robert, B.~Schoenhofer, A.~K. Simonds, and J.~A. Wedzicha.
\newblock Patterns of home mechanical ventilation use in {Europe}: results from
  the {Eurovent} survey.
\newblock {\em European Respiratory Journal}, 25(6):1025--1031, 2005.

\bibitem{hmv_canada2015}
Louise Rose, Douglas~A. McKim, Sherri~L. Katz, David Leasa, Mika Nonoyama,
  Cheryl Pedersen, Roger~S. Goldstein, and Jeremy~D. Road.
\newblock Home mechanical ventilation in {Canada}: A national survey.
\newblock {\em Respiratory Care}, 60(5):695--704, 2015.

\bibitem{who_wheelchair2023}
{World Health Organization}.
\newblock {WHO} releases new wheelchair provision guidelines.
\newblock
  \url{https://www.who.int/news/item/05-06-2023-who-releases-new-wheelchair-provision-guidelines},
  2023.
\newblock WHO estimates that about 80 million people---roughly 1\% of the
  world's population---need a wheelchair. Accessed 2026.

\bibitem{repo}
{bFaaaP Project}.
\newblock {bFaaaP}: Open-source foot-free piano pedal.
\newblock Project site \url{https://bfaaap.com}; repository
  \url{https://github.com/bfaaap/bfaaap_opensource}, 2026.
\newblock Public open-source repository (iOS app, firmware, CAD,
  documentation). Accessed 2026.

\bibitem{videos}
{bFaaaP Project}.
\newblock {bFaaaP}: performance, demonstration, and setup videos.
\newblock YouTube channel
  \url{https://www.youtube.com/channel/UCcAvTy1k8rHs2WrEKPx1amg}.
\newblock Accessed 2026.

\bibitem{concert_kyoen2022}
{Tokyo Tech Platanus and the bFaaaP Project}.
\newblock Natsu-no-kyoen 2022 (a musical festival connecting universities) ---
  ``blumenlied'' (flower song) performed with {bFaaaP}.
\newblock Seven-university joint concert, 2022-09-05, Toyosu Civic Center Hall,
  Tokyo. Video \url{https://youtu.be/uXMLea6_eKM}; event site
  \url{https://natsunokyoen2022.github.io/}, 2022.
\newblock bFaaaP set up among the other performers at a formal multi-university
  concert. Accessed 2026.

\bibitem{concert_suzukake2025}
{Platanus and the bFaaaP Project}.
\newblock 2025 platanus suzukake concert with {bFaaaP} (suzukake science day,
  institute of science tokyo).
\newblock YouTube video \url{https://www.youtube.com/watch?v=V3cXeNW9jXY},
  2025.
\newblock Full grand-piano concert on the {Pro} device; the first author's
  on-instrument setup walkthrough is at 25:01--28:32. Accessed 2026.

\bibitem{video_pro_setup}
{bFaaaP Project}.
\newblock {bFaaaP} {Pro} setup procedure (installation guide).
\newblock YouTube video \url{https://www.youtube.com/watch?v=_9YopbCYTmI},
  2025.
\newblock Step-by-step {Pro} installation: align the drive over the sustain
  pedal, anchor with the airback, set the travel limits. Accessed 2026.

\bibitem{nus}
{Nordic Semiconductor}.
\newblock Nordic {UART} service ({NUS}).
\newblock \url{https://docs.nordicsemi.com/}.
\newblock Accessed 2026.

\bibitem{rp2040}
{Raspberry Pi Ltd.}
\newblock {RP2040} datasheet.
\newblock \url{https://www.raspberrypi.com/documentation/microcontrollers/},
  2024.
\newblock Accessed 2026.

\bibitem{who_great2022}
{World Health Organization and UNICEF}.
\newblock Global report on assistive technology.
\newblock \url{https://www.who.int/publications/i/item/9789240049451}, 2022.
\newblock Reports that $\sim$2.5 billion people need at least one assistive
  product, and that of the $\sim$80 million who need a wheelchair only 5--35\%
  have access depending on country; gives no country-by-country wheelchair
  count. Accessed 2026.

\bibitem{us_census_disability2010}
Matthew~W. Brault.
\newblock Americans with disabilities: 2010.
\newblock Technical Report Current Population Reports P70-131, U.S. Census
  Bureau, 2012.
\newblock About 3.6 million people (1.5\%) aged 15+ used a wheelchair.
  \url{https://www2.census.gov/library/publications/2012/demo/p70-131.pdf}.

\bibitem{nhs_wheelchair}
{NHS England}.
\newblock Wheelchair services.
\newblock \url{https://www.england.nhs.uk/wheelchair-services/}.
\newblock Estimates about 1.2 million wheelchair users in England. Accessed
  2026.

\bibitem{wc_canada_smith2016}
Emma~M. Smith, Edward~M. Giesbrecht, W.~Ben Mortenson, and William~C. Miller.
\newblock Prevalence of wheelchair and scooter use among community-dwelling
  {Canadians}.
\newblock {\em Physical Therapy}, 96(8):1135--1142, 2016.

\bibitem{wc_japan_shirogane2019}
Shun Shirogane et~al.
\newblock Provision of public funding for wheelchairs and postural support
  devices in {Japan}.
\newblock {\em Journal of Physical Therapy Science}, 31(2):122--126, 2019.

\bibitem{wc_australia_aihw}
{Australian Institute of Health and Welfare}.
\newblock People with disability in {Australia}.
\newblock
  \url{https://www.aihw.gov.au/reports/disability/people-with-disability-in-australia},
  2022.
\newblock Drawing on the ABS Survey of Disability, Ageing and Carers (SDAC)
  2018: $\sim$119{,}000 manual-wheelchair users aged 65+, and $\sim$679{,}000
  people with disability use mobility aids. Accessed 2026.

\bibitem{hmv_poland2022}
Ma{\l}gorzata Czajkowska-Malinowska et~al.
\newblock Development of home mechanical ventilation in {Poland} in 2009--2019
  based on the data of the national health fund.
\newblock {\em Journal of Clinical Medicine}, 11(8):2098, 2022.

\bibitem{hmv_hungary2018}
Luca Valk{\'o}, Szabolcs Baglyas, J{\'a}nos G{\'a}l, and Andr{\'a}s Lorx.
\newblock National survey: current prevalence and characteristics of home
  mechanical ventilation in {Hungary}.
\newblock {\em BMC Pulmonary Medicine}, 18:190, 2018.

\bibitem{hmv_korea2019}
H.-I. Kim et~al.
\newblock Home mechanical ventilation use in {South Korea} based on national
  health insurance service data.
\newblock {\em Respiratory Care}, 64(5):528--535, 2019.

\bibitem{hmv_germany2021}
Sarah~B. Schwarz, Maximilian Wollsching-Strobel, Daniel~S. Majorski,
  Friederike~S. Magnet, Tim Mathes, and Wolfram Windisch.
\newblock The development of inpatient initiation and follow-up of home
  mechanical ventilation in {Germany}.
\newblock {\em Deutsches \"Arzteblatt International}, 118(23):403--404, 2021.

\bibitem{us_trach_mehta2015}
Anuj~B. Mehta et~al.
\newblock Trends in tracheostomy for mechanically ventilated patients in the
  {United States}, 1993--2012.
\newblock {\em American Journal of Respiratory and Critical Care Medicine},
  192(4):446--454, 2015.

\bibitem{pct2019}
Tomoyuki Shishido, Hiroyuki Narusawa, Masahiro Ootaki, Kyoko Yamaguchi, and
  Daisuke Tokushige.
\newblock Auxiliary pedal system.
\newblock
  \url{https://patentscope.wipo.int/search/en/detail.jsf?docId=WO2019176164},
  2019.
\newblock PCT International Publication WO~2019/176164~A1 (PCT/JP2018/041771).

\end{thebibliography}

\end{document}